\documentclass[manuscript]{aastex}

\shorttitle{Emergence of the Magnetic Flux tubes}
\shortauthors{Toriumi \& Yokoyama}

\usepackage{subfigure}

\begin{document}


\title{Numerical Experiments on
the Two-step Emergence\\
of Twisted Magnetic Flux Tubes in the Sun}



\author{S. Toriumi and T. Yokoyama}
\affil{Department of Earth and Planetary Science, University of Tokyo,
Hongo, Bunkyo-ku, Tokyo 113-0033, Japan}
\email{toriumi@eps.s.u-tokyo.ac.jp}


\begin{abstract}
We present the new results
of the two-dimensional numerical experiments
on the cross-sectional evolution
of a twisted magnetic flux tube
rising from the deeper solar convection zone
($-20,000\ {\rm km}$)
to the corona through the surface.
The initial depth is ten times deeper
than most of previous calculations
focusing on the flux emergence
from the uppermost convection zone.
We find that the evolution
is illustrated by the two-step process
described below:
the initial tube rises due to its buoyancy,
subject to aerodynamic drag due to
the external flow.
Because of the azimuthal component
of the magnetic field,
the tube maintains its coherency
and does not deform
to become a vortex roll pair.
When the flux tube approaches the photosphere
and expands sufficiently,
the plasma on the rising tube accumulates
to suppress the tube's emergence.
Therefore, the flux decelerates and extends horizontally
beneath the surface.
This new finding owes to
our large scale simulation
calculating simultaneously the dynamics
within the interior
as well as above the surface.
As the magnetic pressure gradient increases
around the surface,
magnetic buoyancy instability
is triggered locally
and, as a result,
the flux rises further into the solar corona.
We also find that the deceleration occurs
at a higher altitude
than in our previous experiment
using magnetic flux sheets
(Toriumi and Yokoyama).
By conducting parametric studies,
we investigate the conditions
for the two-step emergence
of the rising flux tube:
field strength $\ga 1.5\times 10^{4}\ {\rm G}$
and the twist $\ga 5.0\times 10^{-4}\ {\rm km}^{-1}$
at $-20,000\ {\rm km}$ depth.
\end{abstract}

\section{Introduction\label{sec:intro}}

Emerging magnetic fluxes are thought to be
the source of solar active regions
\citep{par55}.
It is widely accepted
that the emerging flux has a form of
an $\Omega$-shaped flux tube.
\citet{par75} studied
the rise time of a flux tube
that emerges due to magnetic buoyancy
under the resistance of aerodynamic drag.
\citet{sch79}
carried out two-dimensional
magnetohydrodynamic (MHD) simulations
and found that the cross-section
of the emerging flux tube
is deformed by the surrounding flows
into an ``umbrella shape''
with a pair of counter-rotating flux rolls,
leading to fragmentation of the tube
\citep[see also][]{lon96}.
After \citet{spr81} introduced
the thin-flux-tube (TFT) approximation,
the emergence of the flux tube has been studied by
using this model
without
considering the effect
of the cross-sectional fragmentation
\citep[e.g.][]{dsi93,fan93,fan94,cal95,mor95}.
\citet{mor96}
and \citet{emo98}
carried out compressible MHD simulations
using the twisted flux tube,
and found that the azimuthal field
can suppress the splitting of the rising tube
into two vortex filaments
\citep[see also][]{dor07}.

Numerical experiments on the flux emergence
above the solar surface have been carried out widely
in the last two decades.
\citet{shi89} produced a pioneering work
on the nonlinear evolution
of the undular mode of magnetic buoyancy instability
(the Parker instability).
The expansion of the magnetic fields into the corona
was found to be in a self-similar way.
\citet{mag01} assumed
a twisted magnetic flux tube
that rises from the uppermost convection zone
to the solar corona.
When the apex of his flux tube
entered the isothermal
(i.e., convectively stable) photosphere,
the tube extended horizontally around the surface.
The rising tube did not develop a wake
with a vortex roll pair,
because the initial tube was located
just beneath the surface ($z=-1800\ {\rm km}$)
and the tube directly entered the photosphere
before the wake is formed.

The transition of the rising flux
through the solar surface
has widely been discussed:
the thorough review in this matter
is found in \citet{mor06}.
\citet{mat92} carried out
the first three-dimensional study
on flux emergence,
and, since then,
various three-dimensional experiments have been done
\citep[e.g.][]{dor99,fan01,arc04,iso05,mur06,man04,gal07,
tor11}.
Recently, MHD calculations including radiative effect
are used to reveal the interaction
between flux emergence and the convective motion
near the surface
\citep{che07,che08,tor09}.
Toroidal flux tubes are also
considered as an initial condition
to study the dynamics of flux emergence
\citep{hoo09,mac09}.

The works introduced above can be
divided into two groups:
one is for the dynamics within the interior
from $\sim -200,000 {\rm km}$ to the surface,
and the other for the emergence
from the uppermost convection zone
(a few $1000\ {\rm km}$ depth)
to the corona.
An additional experiment is required
on the evolution of the emerging flux tube
from the deep convection layer to the corona
through the surface
for the further understanding
of the flux emergence
in a self-consistent manner
\citep{abb03}.
In this paper,
we perform a two-dimensional compressible MHD study
on the buoyant twisted magnetic flux tube
initially embedded deep in the convection zone
($z=-20,000\ {\rm km}$),
of which the depth is the same as our previous
study \citep[hereafter Paper I]{tor10}
and is ten times deeper than
most of previous calculations
considering the emergence
from the uppermost convection zone.

In Paper I,
we carried out numerical experiments
on the emerging flux sheet,
which decelerates in the convection zone
and extends horizontally beneath the surface.
When the magnetic flux accumulates
near the surface
due to the successive emergence below,
the flux sheet is subject to
the Parker instability
and thus further evolution to the corona occurs
(the ``two-step emergence'' model).
When the field is too weak,
the rising flux cannot evolve
further to the corona
and remains within the solar interior
(``failed emergence''),
while, in the case of a strong field,
the flux directly emerges
to the upper atmosphere
without showing a deceleration
(``direct emergence'').

The results of the present study
are similar to the previous experiments.
In the convection zone,
the cross-sectional evolution of the flux tube
is similar to the papers by
\citet{mor96} and \citet{emo98}.
We found that,
as the flux tube rises,
aerodynamic drag becomes more effective,
since the external flow from the apex
forms a wake behind the main tube.
Also,
the emergence above the surface
is similar to the calculations
by \citet{mag01}.
However, there is an important difference
from those papers;
the deceleration of the rising tube
occurs when the tube is far below the photosphere,
and the tube makes a flattened $\nabla$-shaped structure
widely beneath the surface.
By conducting an analytic model
which includes the effect of the mass pile-up,
we confirm that
the tube's deceleration
is the consequence of the fluid accumulation
between the rising tube and the photosphere.
The magnetic flux stretched horizontally
beneath the surface
becomes magnetic buoyancy-unstable locally
so that the second-step emergence
to the corona takes place.
These features resemble the previous two-step model in Paper I.
We also carry out parameter surveys
to study the dependence
on the initial twist
and the field strength.

In the next section (\S \ref{sec:model}),
we give a description of the model
used in this study,
while the results of the numerical experiments
are shown in detail
in Section \ref{sec:results}.
In Section \ref{sec:param},
we present the results of the parameter study.
Finally, in Section \ref{sec:discuss},
we summarize and discuss the results.

\section{Numerical Model\label{sec:model}}

\subsection{Assumptions and Basic Equations}
We consider the buoyant rise of 
an isolated magnetic flux tube 
in the stratified ideal gas layers
in the $(y,\,z)$-plane,
where the $z$-coordinate increases upward.
We solve the standard set of ideal MHD equations
including constant gravitational acceleration
$\mbox{\boldmath $g$}=(0,0,-g_{0})$.
The basic equations are the same as
those of Paper I:
\begin{eqnarray}
	\frac{\partial\rho}{\partial t}
	 + \mbox{\boldmath $\nabla$}
	 \cdot(\rho \mbox{\boldmath $V$})=0\ ,
\end{eqnarray}
\begin{eqnarray}
	\frac{\partial(\rho \mbox{\boldmath $V$})}{\partial t}
	+ \mbox{\boldmath $\nabla$}\cdot
	\left(
	 \rho \mbox{\boldmath $VV$}+p\mbox{\boldmath $I$} 
	 -\frac{\mbox{\boldmath $BB$}}{4\pi}
	 +\frac{\mbox{\boldmath $B$}^{2}}{8\pi}\mbox{\boldmath $I$}
	\right)
	-\rho \mbox{\boldmath $g$} =0\ ,
\end{eqnarray}
\begin{eqnarray}
	\frac{\partial \mbox{\boldmath $B$}}{\partial t}
	= \mbox{\boldmath $\nabla$}\times
	(\mbox{\boldmath $V$}\times \mbox{\boldmath $B$})\ ,
\end{eqnarray}
\begin{eqnarray}
	\frac{\partial}{\partial t}
	\left(
	 \rho U+\frac{1}{2}\rho \mbox{\boldmath $V$}^{2}
	 +\frac{\mbox{\boldmath $B$}^{2}}{8\pi}
	\right)
	+\mbox{\boldmath $\nabla$}\cdot
	\left[
	 (\rho U+p+\frac{1}{2}\rho \mbox{\boldmath $V$}^{2})
	 \mbox{\boldmath $V$}
	 +\frac{c}{4\pi}\mbox{\boldmath $E$}\times \mbox{\boldmath $B$}
	\right]
	-\rho \mbox{\boldmath $g$}\cdot \mbox{\boldmath $V$}=0\ ,
\end{eqnarray}
and
\begin{eqnarray}
	U=\frac{1}{\gamma -1}\frac{p}{\rho}\ ,
\end{eqnarray}
\begin{eqnarray}
	\mbox{\boldmath $E$}
	 =-\frac{1}{c}\mbox{\boldmath $V$}
          \times \mbox{\boldmath $B$}\ ,
\end{eqnarray}
\begin{eqnarray}
        p=\frac{k_{\rm B}}{m}\rho T\ ,
\end{eqnarray}
where $U$ is the internal energy per unit mass,
$\mbox{\boldmath $I$}$ is the unit tensor,
$k_{\rm B}$ is the Boltzmann constant,
$m$ is the mean molecular mass,
and other symbols have their usual meanings.
In this study,
we carry out so-called 2.5-dimensional numerical simulation,
that is, all the physical quantities are
independent of $x$
while the $x$-component of vector quantities
(i.e., velocity \mbox{\boldmath $V$}
and magnetic field \mbox{\boldmath $B$})
is taken into account.
We assume a ratio of specific heats, $\gamma =5/3$.

The normalizing units of length, velocity, time, and density
in the simulations
are $H_{0}$, $C_{s0}$, $\tau_{0}\equiv H_{0}/C_{s0}$,
and $\rho_{0}$, respectively,
where $H_{0}=k_{\rm B}T_{0}/(mg_{0})$ is
the pressure scale height,
$C_{s0}$ the sound speed,
and $\rho_{0}$ the density at the photosphere.
The gas pressure, temperature, and magnetic field strength
are normalized by the combinations of the units above,
i.e., $p_{0}=\rho_{0}C_{s0}^{2}$,
$T_{0}=mC_{s0}^{2}/(\gamma k_{\rm B})$,
and $B_{0}=(\rho_{0}C_{s0}^{2})^{1/2}$, respectively.
The gravity is given as $g_{0}=C_{s0}^{2}/(\gamma H_{0})$
by definition.
For the comparison of numerical results with observations,
we use $H_{0}=200\ {\rm km}$, $C_{s0}=8\ {\rm km\,s}^{-1}$,
$\tau_{0}=H_{0}/C_{s0}=25\ {\rm s}$,
and $\rho_{0}=1.4\times 10^{-7}\ {\rm g\ cm}^{-3}$,
which are typical values
for the solar photosphere and chromosphere.
Then, $p_{0}=9.0\times 10^{4}\ {\rm dyn\ cm}^{-2}$,
$T_{0}=4000\ {\rm K}$, and $B_{0}=300\,{\rm G}$.

\subsection{Initial Conditions}
The initial background stratification
consists of three regions
\citep[see e.g.][]{noz92}:
an adiabatically stratified convection zone,
a cool isothermal photosphere/chromosphere,
and a hot isothermal solar corona.
We take $z=0$ to be the base height of the photosphere,
and the initial temperature distribution
of the photosphere/chromosphere and the corona ($z\geq 0$)
is assumed to be
\begin{eqnarray}
  T = T_{\rm s}(z)
    \equiv T_{\rm ph} + (T_{\rm cor} - T_{\rm ph})
         \{\tanh{[(z-z_{\rm cor})/w_{\rm tr}]}+1\}/2\ ,
  \label{eq:temp1}
\end{eqnarray}
where $T_{\rm ph}=T_{0}$ and $T_{\rm cor}=100T_{0}$ are
the respective temperatures
in the photosphere/chromosphere and the corona,
$z_{\rm cor}=10H_{0}$ is the base of the corona,
and $w_{\rm tr}=0.5H_{0}$ is
the temperature scale height of the transition region.
The initial temperature distribution
in the convection zone ($z\leq 0$) is described as
\begin{eqnarray}
  T = T_{\rm s}(z) \equiv T_{\rm ph} -
  z \left|
     \frac{dT}{dz}
    \right|_{\rm ad}\ ,
  \label{eq:temp2}
\end{eqnarray}
where
\begin{eqnarray}
  \left|
   \frac{dT}{dz}
  \right|_{\rm ad}
  =\frac{\gamma -1}{\gamma}\frac{mg_{0}}{k_{\rm B}}
\end{eqnarray}
is the adiabatic temperature gradient,
i.e., the initial temperature distribution
in the convection zone is adiabatic.
On the basis of the temperature distribution above,
the initial pressure and density profiles
are defined by the equation of static pressure balance:
\begin{eqnarray}
  \frac{dp_{\rm s}(z)}{dz}+\rho_{\rm s}(z) g_{0}=0 .
  \label{eq:pre}
\end{eqnarray}

\begin{figure}
\epsscale{0.8}
\plotone{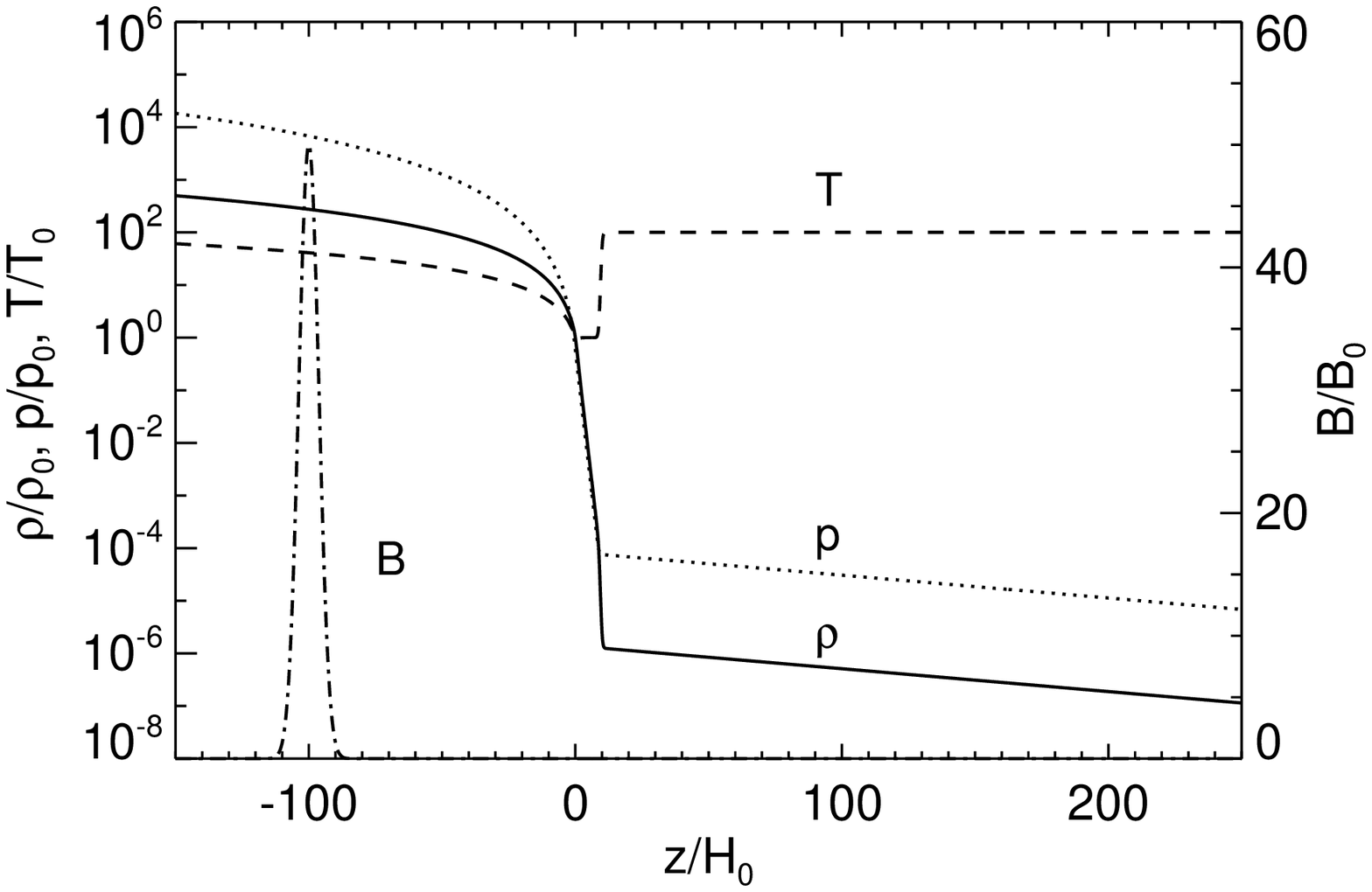}
\caption{One-dimensional ($z$-)distributions
 of the initial background density (solid line),
 pressure (dotted line), and temperature (dashed line).
 The total magnetic field strength
 $B=[B_{x}^2+B_{\phi}^2]^{1/2}$
 of case 1
 along the vertical axis $y=0$
 is overplotted with a dashed-dotted line.}
\label{fig:ini}
\end{figure}

The initial magnetic flux tube is embedded
in the solar interior at $z=-100H_{0}=-20,000\ {\rm km}$.
The longitudinal and azimuthal components
of the flux tube
are described as follows
\citep[see][]{fan01,arc04,mur06}:
for $r\equiv [(y-y_{\rm tube})^{2}+(z-z_{\rm tube})^{2}]^{1/2}$,
\begin{eqnarray}
  B_{x}(r) = B_{\rm tube}
  \exp{\left( -\frac{r^{2}}{R_{\rm tube}^2} \right)}\ ,
\end{eqnarray}
and
\begin{eqnarray}
  B_{\phi}(r) = qr B_{x}(r)\ ,
\end{eqnarray}
where $(y_{\rm tube},\ z_{\rm tube})=(0,\ -100H_{0})$
is the tube center,
$R_{\rm tube}$ is the tube radius,
$q$ is the twist parameter
denoting the angular rate of field lines
rotating around the tube's axis
per unit length,
and $B_{\rm tube}$ is the field strength at the center.
The horizontal and vertical components
of the azimuthal field are defined as
\begin{eqnarray}
  B_{y}(y,\,z) = -B_{\phi}(r) \frac{z-z_{\rm tube}}{r}\ ,
\end{eqnarray}
and
\begin{eqnarray}
  B_{z}(y,\,z) = B_{\phi}(r) \frac{y-y_{\rm tube}}{r}\ ,
\end{eqnarray}
respectively.
We use $R_{\rm tube}=5H_{0}=1000\ {\rm km}$
throughout the paper.
For the typical case (case 1),
we take $B_{\rm tube}=50B_{0}=1.5\times 10^{4}\ {\rm G}$,
so the total longitudinal magnetic flux is
$\Phi_{x}
%=\int_{0}^{R_{\rm tube}}B_{x}(r)\,2\pi r\,dr
=4.7\times 10^{20}\ {\rm Mx}$.
The twist parameter is $q=0.1/H_{0}$ for case 1.
The pressure inside the tube is defined as
\begin{eqnarray}
  p_{\rm i}(y,\,z) = p_{\rm s}(z) + \delta p_{\rm exc}(r)\ ,
\end{eqnarray}
where $\delta p_{\rm exc}(<0)$ is the pressure excess
described as
\begin{eqnarray}
  \delta p_{\rm exc}(r)
  = \frac{B_{x}^{2}(r)}{8\pi}
  \left[
    q^{2}
    \left(
     \frac{R_{\rm tube}^{2}}{2} - r^{2}
    \right)
    -1
  \right]\ ,
\end{eqnarray}
for the pressure balance.
The temperature is kept unchanged,
i.e., thermal balance is sustained,
$T_{\rm i}(z)=T_{\rm s}(z)$.
Since the density inside the tube
is smaller than that outside,
the flux tube will buoyantly rise through the convection zone.
The initial temperature, density, and pressure profiles
of the background stratification,
and the total field strength
$B=[B_{x}^{2}+B_{\phi}^{2}]^{1/2}
=[B_{x}^{2}+B_{y}^{2}+B_{z}^{2}]^{1/2}$
of case 1 along $y=0$
are shown in Figure \ref{fig:ini}.

\subsection{Boundary Conditions and Numerical Procedures}
The domain of the simulation box is
$(y_{\rm min}<y<y_{\rm max})$
and $(z_{\rm min}<z<z_{\rm max})$,
where $y_{\rm min}=-400H_{0}$, $y_{\rm max}=400H_{0}$,
$z_{\rm min}=-150H_{0}$, and $z_{\rm max}=250H_{0}$.
The total size of the simulation box is
$160\ {\rm Mm}\times 80\ {\rm Mm}$.
Periodic boundaries are assumed for
$y=y_{\rm min}$ and $y=y_{\rm max}$,
symmetric boundaries for
$z=z_{\rm min}$ and $z=z_{\rm max}$.
A wave-damping region is attached near the top boundary.
The total number of grid points is
$(N_{y}\times N_{z})=(1536\times 1920)$,
and the mesh sizes are
$\Delta y=0.52H_{0}$ and $\Delta z=0.21H_{0}$,
both of which are uniform.
We use the modified Lax-Wendroff scheme
version of the CANS (Coordinated Astronomical Numerical Software)
code (see Paper I).
To study the dependence on the twist
and the field strength of the flux tube,
we change the values for the twist parameter $q$
or the axial field strength $B_{\rm tube}$.
The cases we use in this study are summarized
in Table \ref{tab:param}.

\section{Results\label{sec:results}}

Here, we show the results
of the typical model (case 1)
in which the initial twisted flux tube with $q=0.1/H_{0}$
rises through the convection zone to the corona.
The evolution within the interior
resembles the studies by \citet{mor96} and \citet{emo98},
and the evolution to the upper atmosphere
is similar to \citet{mag01}.
However, the present result is not a simple sum of them;
the rising tube decelerates within the interior
far below the surface
to build a widely flattened magnetic structure.

\begin{figure}
\begin{center}
\includegraphics[clip,scale=0.45]{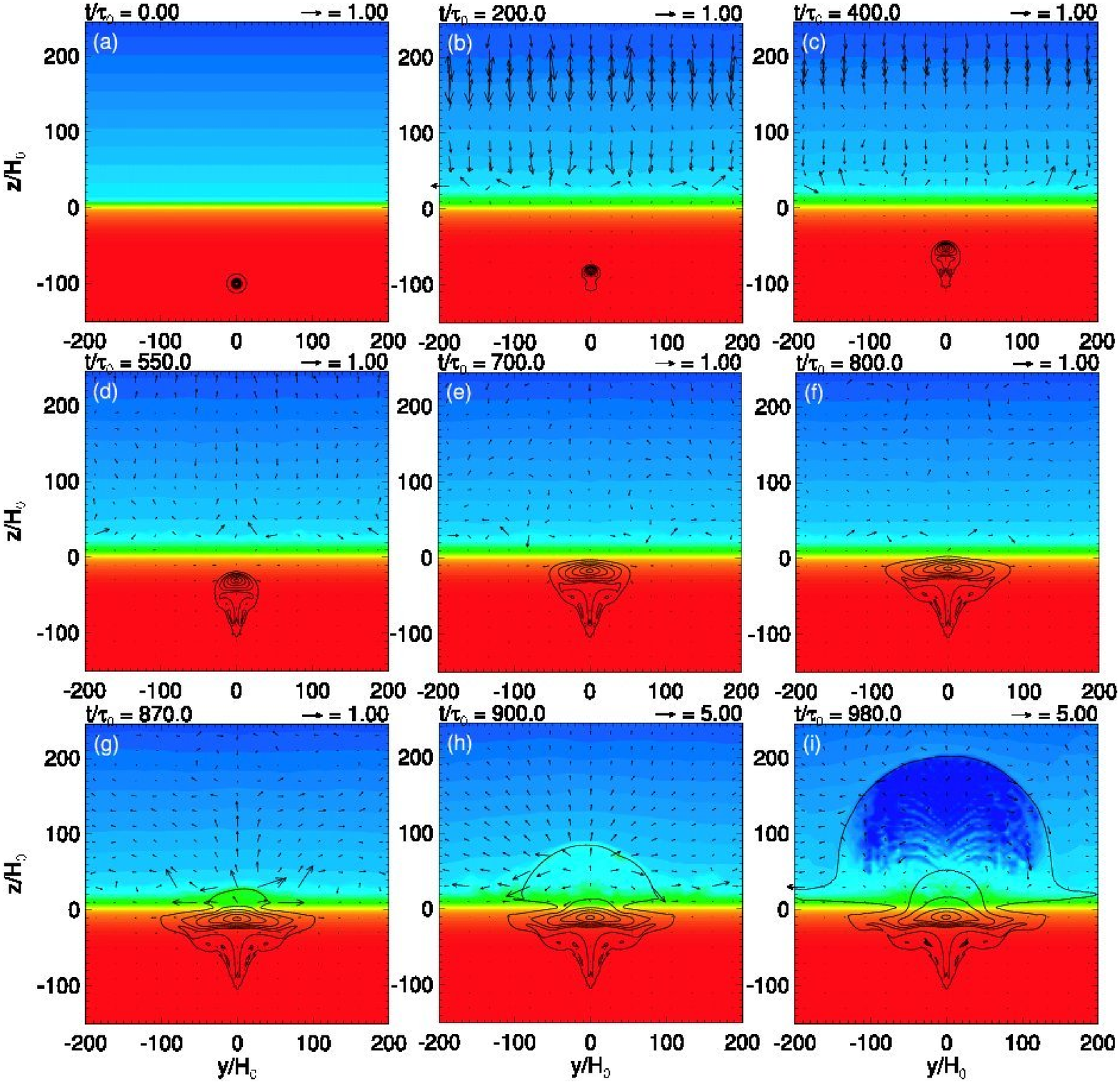}\\
\includegraphics[clip,scale=0.6,angle=-90.]{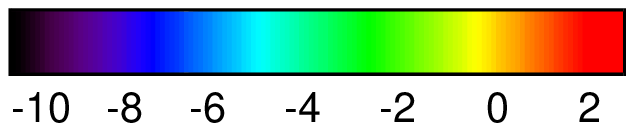}
\caption{Time-evolution of the ``two-step emergence''
of the twisted flux tube
(typical model).
(a) $t/\tau_{0}=0$;
(b) $t/\tau_{0}=200$;
(c) $t/\tau_{0}=400$;
(d) $t/\tau_{0}=550$;
(e) $t/\tau_{0}=700$;
(f) $t/\tau_{0}=800$;
(g) $t/\tau_{0}=870$;
(h) $t/\tau_{0}=900$;
(i) $t/\tau_{0}=980$.
Logarithmic density profiles ($\log_{10}{(\rho/\rho_{0})}$)
are indicated by color contours,
while magnetic field lines and velocity vectors are overplotted with
black lines and arrows, respectively.
This figure is also available as an avi animation in the electronic
edition of the {\it Astrophysical Journal}.}
\label{fig:ro}
\end{center}
\end{figure}

\begin{figure}
\begin{center}
\subfigure{\includegraphics[clip,scale=0.5]{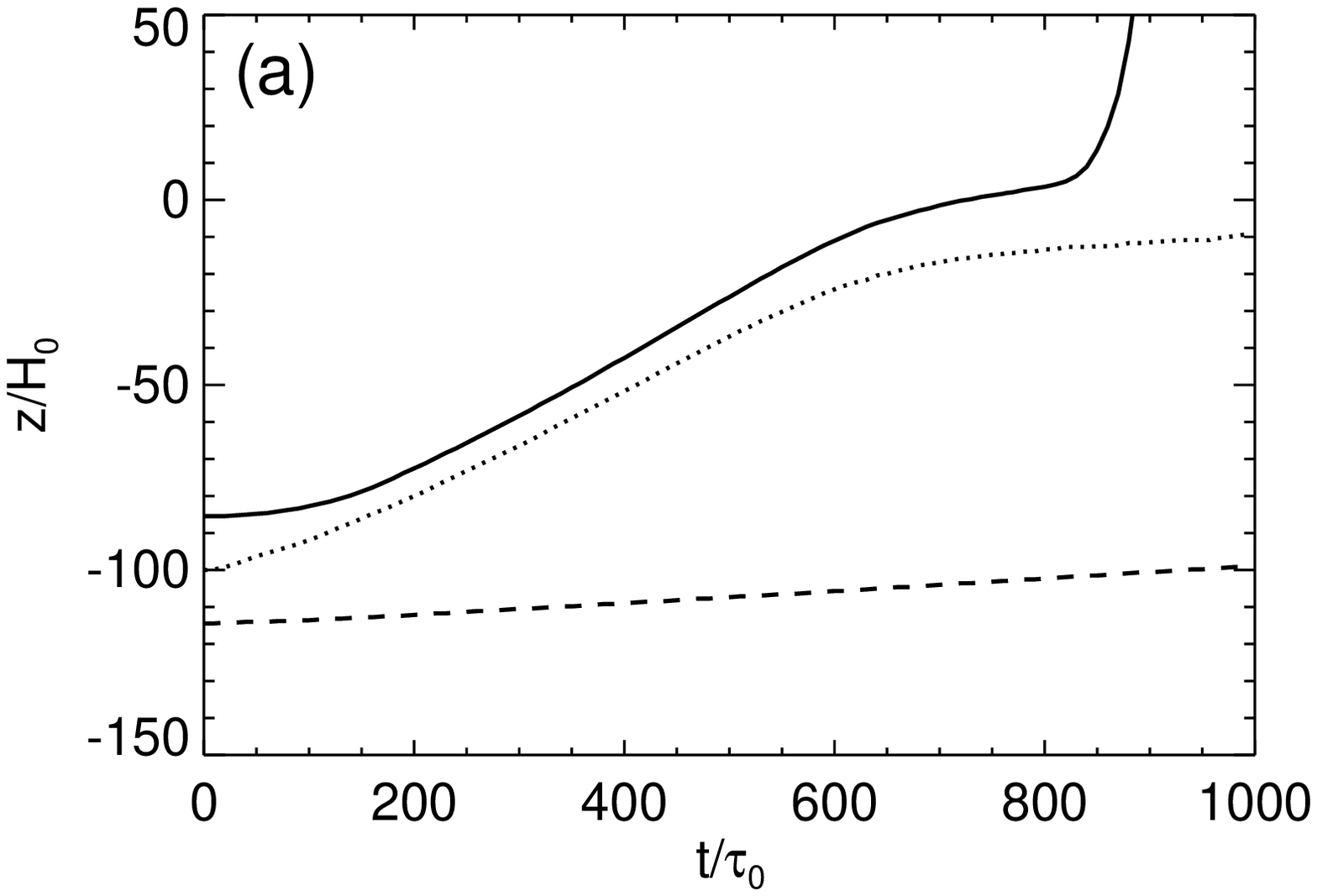}
\label{fig:z_3}}\\
\subfigure{\includegraphics[clip,scale=0.5]{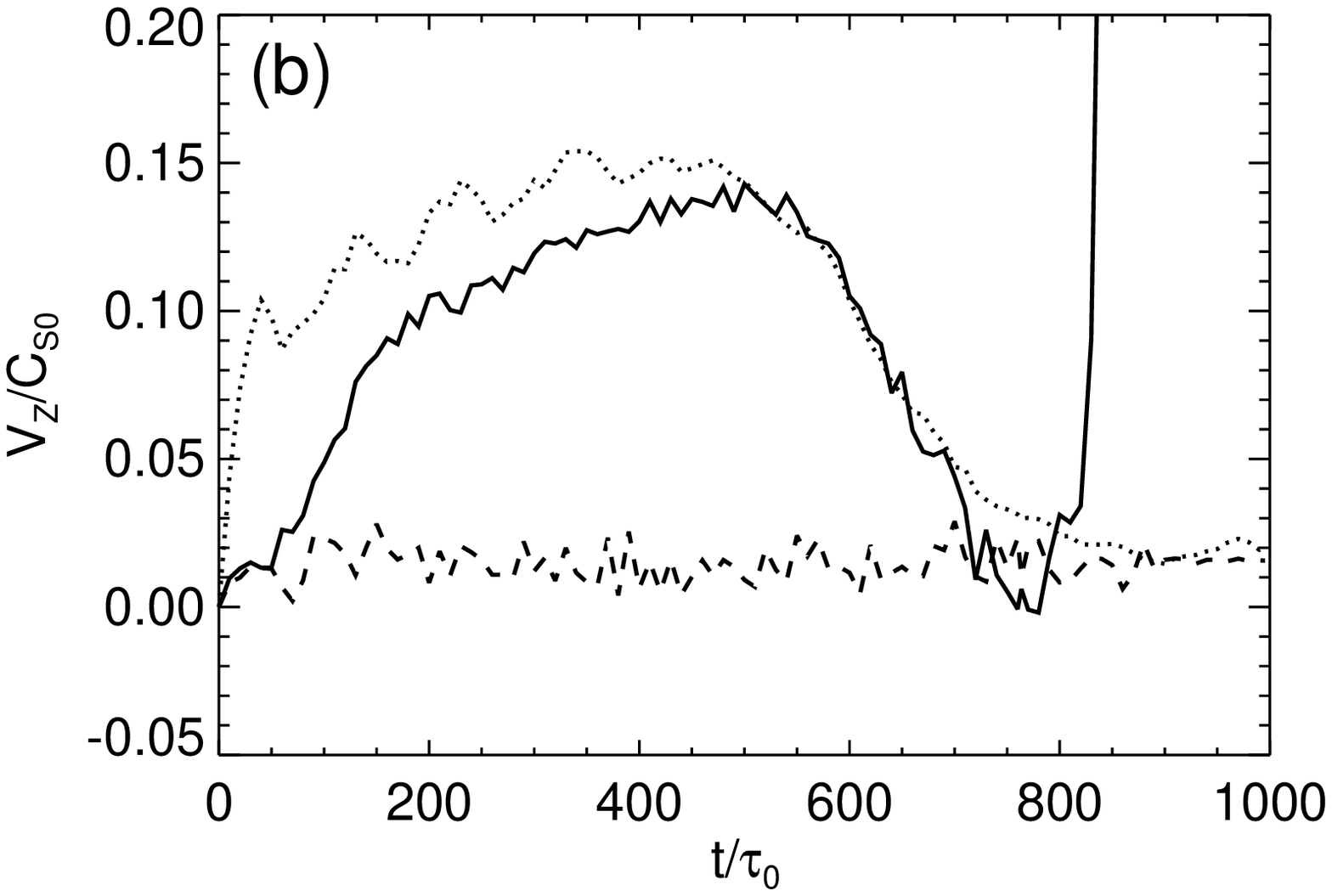}
\label{fig:vz_3}}
\caption{\subref{fig:z_3}: Height-time relations
at the top (solid line), the center (dotted line),
and the bottom (dashed line)
of the flux tube.
\subref{fig:vz_3}: Gas velocities at these three points.}
\label{fig:z_vz}
\end{center}
\end{figure}

Figure \ref{fig:ro} displays
the evolution of the density profile (color contour)
in a vertical plane normal to the tube axis,
while the solid lines and arrows
indicate the magnetic field lines and the velocity vectors,
respectively.
Figure \ref{fig:ro}(a) presents the initial state;
the flux tube is located at $z/H_{0}=-100$
with a circular shape,
and begins to emerge because of the buoyancy
due to its smaller density
relative to the surroundings.
Figure \ref{fig:z_3} shows
the height-time relations
at the top (solid line),
the center (dotted line),
and the bottom (dashed line)
of the emerging tube,
while the velocities at these points
are presented in Figure \ref{fig:vz_3}.
In the initial phase ($0<t/\tau_{0}<100$),
the flux tube rises due to magnetic buoyancy
and the tube's cross-section
almost keeps its original circular shape.
Thus, the tube
moves up with a constant acceleration rate.
In the next phase ($100<t/\tau_{0}<500$),
the aerodynamic drag
grows to counteract buoyancy,
since the external flow
around the tube's cross-section
develops a wake
behind the main tube
(Figure \ref{fig:ro}(c)),
and the acceleration of the rise velocity
is reduced.
The azimuthal field of the flux tube resists
this deformation
and prevents the tube from fragmenting
into a counter-rotating vortex pair
(the so-called ``umbrella shape''
in \citet{sch79}).
During the emergence in the convection zone,
the rise velocity of the tube center fluctuates
due to the internal torsional oscillation
(Figure \ref{fig:vz_3}; see \S \ref{sec:conv}).
As the flux tube reaches close to the photosphere,
it expands horizontally
(Figure \ref{fig:ro}(e))
and its rise speed decreases
after $t/\tau_{0}=500$.
Eventually, the subsurface flux
extends widely in the range of $-100<y/H_{0}<100$
at $t/\tau_{0}=800$
(Figure \ref{fig:ro}(f)).
When the top of the subphotospheric flux
becomes unstable to the magnetic buoyancy instability,
further evolution to the corona breaks out
(Figures \ref{fig:ro}(f)-(i);
see \S \ref{sec:corona}).
The consequent coronal loop
at the time $t/\tau_{0}=980$
has its width $\sim 400H_{0}=80,000\ {\rm km}$
and height $\sim 200H_{0}=40,000\ {\rm km}$.
The picture of this evolution
resembles our previous ``two-step emergence'' model
(Paper I).
In the following sections,
we will describe some more detailed results,
especially on the magnetic and the flow field
within the convection zone,
the comparison with the analytic model,
and the second-step evolution starting at the photosphere.

\subsection{Magnetic Fields and Vorticity\label{sec:conv}}
\begin{figure}
\begin{center}
\includegraphics[clip,scale=0.4]{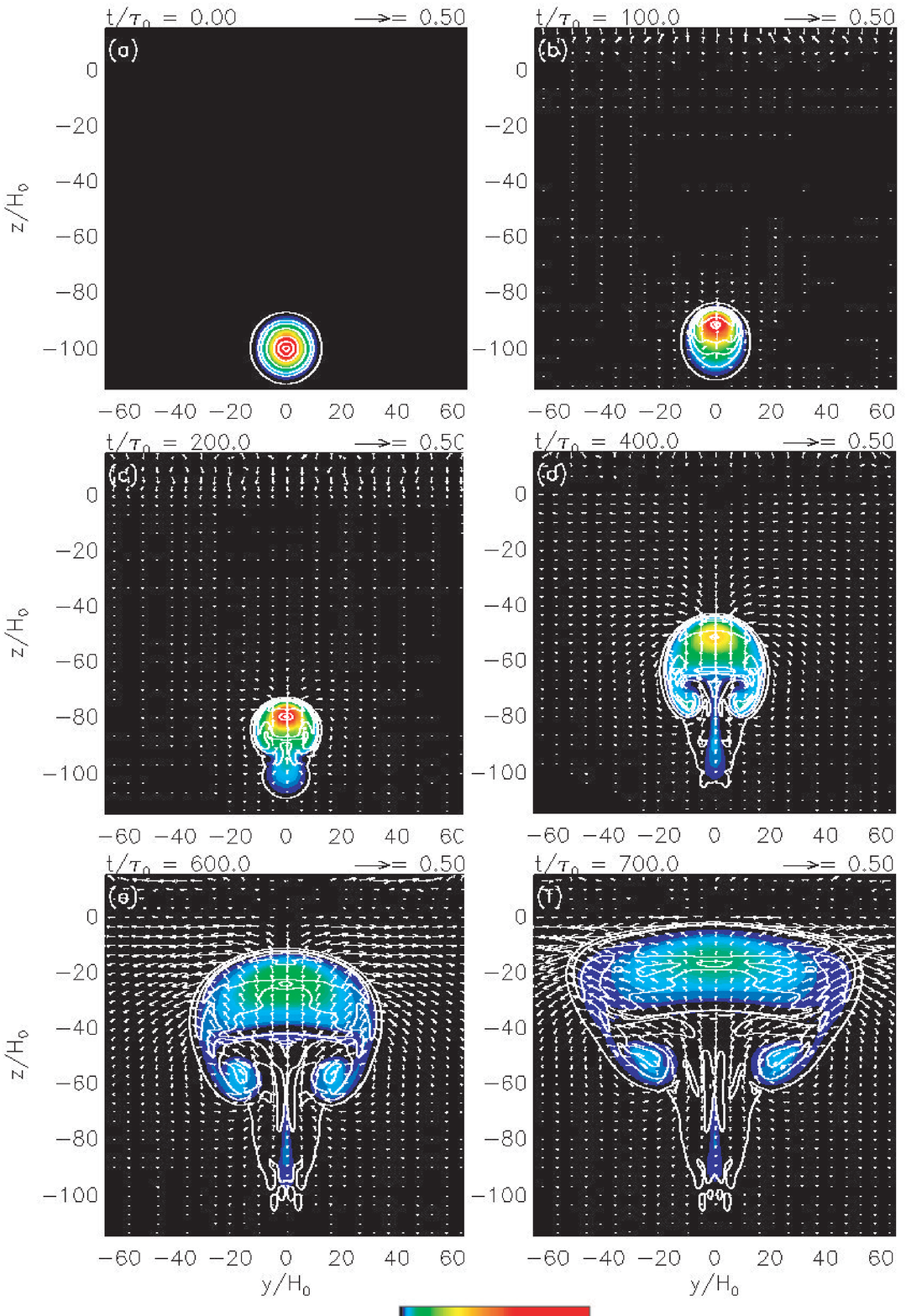}\\
\includegraphics[clip,scale=0.6,angle=-90.]{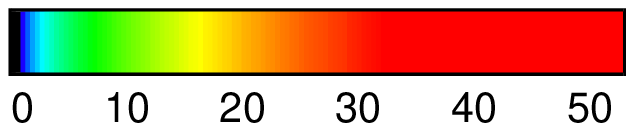}
\caption{Time-evolution of the magnetic fields
from $t/\tau_{0}=0$ to $t/\tau_{0}=700$.
Color contour displays the longitudinal component
$B_{x}/B_{0}$,
while the azimuthal component $B_{\phi}/B_{0}$
is overplotted with solid lines.}
\label{fig:mag}
\end{center}
\end{figure}

\begin{figure}
\begin{center}
\includegraphics[clip,scale=0.4]{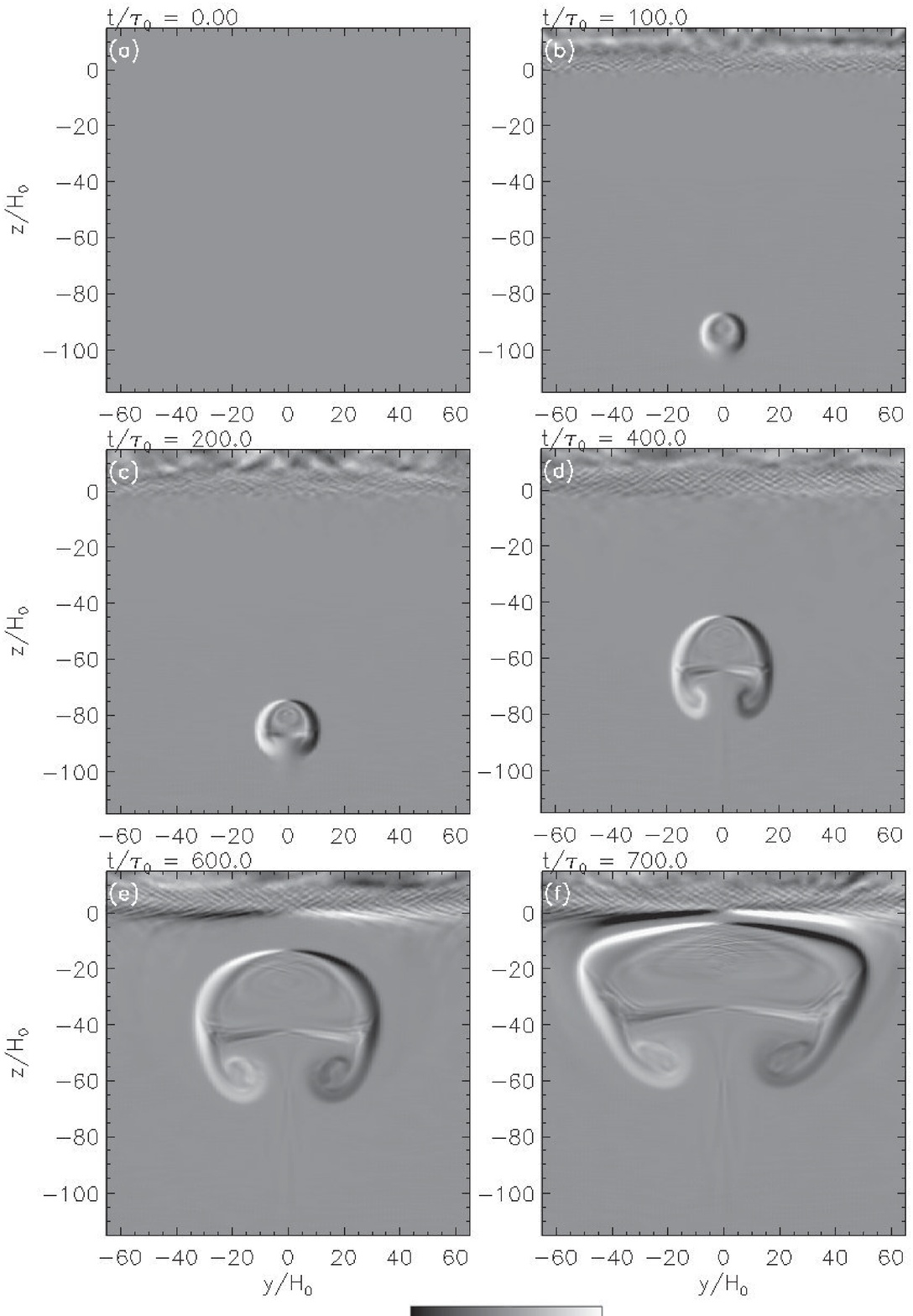}\\
\includegraphics[clip,scale=0.6,angle=-90.]{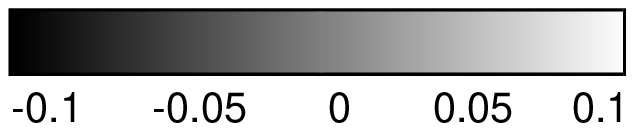}
\caption{Same as Figure \ref{fig:mag}
but for the $x$-component of the vorticity.
White indicates positive (counter-clockwise) vorticity,
while black is negative (clockwise).
}
\label{fig:vor}
\end{center}
\end{figure}

To study the flux tube's emergence in the solar interior,
we show the variation of the field configuration with time
in Figure \ref{fig:mag}.
The color contour indicates
the distribution of the longitudinal field ($B_{x}$)
while the strength of the azimuthal field ($B_{\phi}$)
is overplotted with solid lines.
In Figure \ref{fig:vor},
we plot the $x$-component of the vorticity vector
$(\omega_{x})$,
where
\begin{eqnarray}
  \omega_{x}=\frac{\partial V_{z}}{\partial y}
            -\frac{\partial V_{y}}{\partial z}\ ,
\end{eqnarray}
and white and black colors indicate
positive (counter-clockwise)
and negative vorticity (clockwise), respectively.
These figures (Figures \ref{fig:mag} and \ref{fig:vor})
look similar to those of previous experiments
within the convection zone
\citep{mor96,emo98}.
Figures \ref{fig:mag}(a) and \ref{fig:vor}(a)
present the initial state:
the flux tube has a circular shape
and there is no vorticity within the domain.
As the tube begins to emerge,
the field strength reduces and the tube expands,
since the background density decreases with height
(Figures \ref{fig:mag}(b)-(d)).
Also, the external flow
from the apex to the flanks of the tube
creates negative and positive vortex sheets
in the right and left side
of the main tube,
respectively
(Figures \ref{fig:vor}(b)-(d)).
Therefore, the tube is deformed
and the wake behind the main tube grows
with a pair of counter-rotating vortex rolls.
Because of the azimuthal fields, however,
the tube is not entirely fragmented
into an ``umbrella shape.''
As a consequence,
the rising tube feels aerodynamic drag,
and, thus,
the rise velocity levels off
at this time.
The longitudinal field component keeps its strength
in the core of the vortex rolls
as well as in the main tube
(see Figure \ref{fig:mag}(e)).
Behind the main tube,
in $-100<z/H_{0}<-60$ in Figure \ref{fig:mag}(f),
there is a weakly magnetized tail
with non-zero vorticity (Figure \ref{fig:vor}(f)).

During this phase,
the rise speed of the tube center
reveals oscillation as seen in Figure \ref{fig:vz_3}.
This is a result of torsional oscillation
due to the differential buoyancy
caused by the magnetic field distribution
\citep[e.g.][]{mor96}.
Initially, the flux tube
has its maximum field strength at the center
and thus the central region rises faster than the periphery.
Therefore, the azimuthal field around the apex
is compressed and strengthened
to fortify the magnetic tension force,
which inhibits the tube from a distortion
into vortex rolls.
The tension force decelerates the center of the tube
and, as a result,
induces the internal torsional oscillation.
The period of this oscillation
is observed to be consistent with
the time of the azimuthal Alfv$\acute{\rm e}$n speed
traveling across the tube's diameter.

At around $t/\tau_{0}=500$,
the rise motion begins to decelerate
and the front of the tube expands horizontally,
since the tube is close enough to the surface.
The plasma above the rising tube
cannot move through the isothermal
(i.e., convectively stable) photosphere
so that the fluid is compressed and piles up
between the front of the tube and the photosphere.
The fluid between them suppresses
the rising motion of the flux below,
and thus, the flux extends sideways.
The horizontal expansion
caused by the photosphere
was previously reported
by \citet{mag01}.
His flux tube was not so deformed
as to develop a wake including vortex rolls
and directly entered the photosphere
mostly keeping its original cylindrical shape,
because the initial tube was located
just beneath the surface
($z_{\rm tube}= -1800\ {\rm km}$)
with a strong field strength
($B_{\rm tube}= 7760\ {\rm G}$).
In our case,
the initial tube was embedded
so deep in the convection zone
($z_{\rm tube}= -20,000\ {\rm km}$)
that the wake
behind the main tube was formed
and the drag force became more effective.
The velocity shear
between the horizontal flow above the tube
and the photosphere
forms vortex sheets at around $z/H_{0}\sim 0$
(Figure \ref{fig:vor}(e);
the positive sheet in the right half,
and the negative in the left).
Finally, at $t/\tau_{0}=700$,
the flux tube has a $\nabla$-shaped structure
and the magnetic field extends
widely beneath the surface
(Figure \ref{fig:mag}(f)).
The vortex sheet at the tube front
collides against the subphotospheric sheet
(Figure \ref{fig:vor}(f)).
In this stage,
the azimuthal field beneath the surface
is much stronger than the longitudinal field
because of the radial expansion of the flux tube
\citep{par74,par79}.

\subsection{Comparison with an Analytic Model
\label{sec:analytic}}
\cite{fan98} compared
their numerical results
of the emerging twisted flux tube
obtained by 2.5D anelastic MHD experiments
with those of the thin-flux-tube model.
According to \cite{fan98},
the motion of the cylindrical flux tube
rising by its magnetic buoyancy is described as
\begin{eqnarray}
  I\rho \frac{dV_{z}}{dt}
  = -\Delta\rho g - C_{\rm D} \frac{\rho|V_{z}|V_{z}}{\pi R}\ ,
  \label{eq:eqm1}
\end{eqnarray}
where $I$ is the enhanced inertia factor ($\sim 2$),
$C_{\rm D}$ is the drag co-efficient of order unity,
and $\Delta\rho=\rho_{\rm i}-\rho$
is the density difference
between the flux tube ($\rho_{\rm i}$)
and the external medium ($\rho$).
From Equation (\ref{eq:eqm1}),
one can calculate the rising velocity
of the tube $V_{z}(t)$
and the height of the tube center $z(t)$.
We consider the effect of the mass pile-up
between the flux tube and the photosphere
and replace $\Delta\rho$
with an additional effect as
\begin{eqnarray}
  \Delta\rho'&=&\Delta\rho+F\Delta\rho_{\rm acm}\nonumber\\
             &=&\rho_{\rm i}-\rho
        + F\left[\bar{\rho}_{\rm acm}(0)
                 -\bar{\rho}_{\rm acm}(t)\right]\ .
\end{eqnarray}
Here,
\begin{eqnarray}
  \bar{\rho}_{\rm acm}(0)
  =\frac{1}{z_{\rm ph}-z(0)}
   \int_{z(0)}^{z_{\rm ph}}\rho_{\rm s}(\zeta)\,d\zeta
\end{eqnarray}
is the plasma on the flux tube in the initial state
($z_{\rm ph}=0$, $z(0)=z_{\rm tube}=-100H_{0}$,
and $\rho_{\rm s}(z)$ is given by Equation (\ref{eq:pre})),
and
\begin{eqnarray}
  \bar{\rho}_{\rm acm}(t)
  =\frac{1}{z_{\rm ph}-z(t)}
   \int_{z(t)}^{z_{\rm ph}}\rho_{\rm s}(\zeta)\,d\zeta
\end{eqnarray}
is the background plasma at a given time.
That is,
$\Delta\rho_{\rm acm}
=\bar{\rho}_{\rm acm}(0)-\bar{\rho}_{\rm acm}(t)$
corresponds to a density
that would have been accumulated
by an ideal rising sheet
extending horizontally.
To consider the effect of the tube's shape
and the draining of the plasma from the apex,
we multiply it by a factor $F(<1)$,
which
depends on the field geometry
and its initial depth from the surface
($R_{\rm tube}$ and $z_{\rm tube}$ ).
In the present study,
we vary $F$ to fit the analytic model
to the obtained data
and assume it as a constant.
The relation between $F$
and $R_{\rm tube}$ and $z_{\rm tube}$
requires much work,
which we shall leave for future research.
Applying $\Delta\rho'$
and dividing Equation (\ref{eq:eqm1}) by $\rho$,
we get
\begin{eqnarray}
  I\frac{dV_{z}}{dt}
  = -\frac{\Delta\rho'}{\rho} g
    - \frac{C_{\rm D}}{\pi R} |V_{z}|V_{z}\ .
  \label{eq:eqm2}
\end{eqnarray}
If pressure balance, mass and flux conservation,
and adiabatic evolution are assumed,
the buoyancy and the radius of the model tube
can be defined as
\begin{eqnarray}
  -\frac{\Delta\rho'}{\rho}g
  = \left(
      -\frac{\Delta\rho}{\rho}g
    \right)_{z=z_{\rm tube}}
    \left[
      \frac{(\Gamma+1)-z/H_{0}}{(\Gamma+1)-z_{\rm tube}/H_{0}}
    \right]^{\Gamma-1}
    +F\frac{\Delta\rho_{\rm acm}}{\rho}\ ,
\end{eqnarray}
\begin{eqnarray}
  R(z) = R(z=z_{\rm tube})
    \left[
      \frac{(\Gamma+1)-z/H_{0}}{(\Gamma+1)-z_{\rm tube}/H_{0}}
    \right]^{\Gamma/2}\ ,
\end{eqnarray}
where $\Gamma=1/(\gamma -1)$.
Here we use their initial values
obtained as
\begin{eqnarray}
  \left(
    -\frac{\Delta\rho}{\rho}g
  \right)_{z=z_{\rm tube}}
  = \frac{
      \int\!\!\!\int (-\Delta\rho/\rho)g\,B_{x}\,dydz
    }{
      \int\!\!\!\int B_{x}\,dydz
    }\ ,
  \label{eq:ini1}
\end{eqnarray}
and
\begin{eqnarray}
  R(z=z_{\rm tube})
  = \left[
    \frac{
      \int\!\!\!\int [(y-y_{\rm tube})^{2}
	        +(z-z_{\rm tube})^{2}]
      \,B_{x}\,dydz
    }{
      \int\!\!\!\int B_{x}\,dydz
    }
    \right]^{1/2}
  \label{eq:ini2}
\end{eqnarray}
\citep[see][]{fan98}.
We can calculate the time evolution of the model tube
by integrating Equation (\ref{eq:eqm2}).

\begin{figure}
\begin{center}
\subfigure{\includegraphics[clip,scale=0.5]{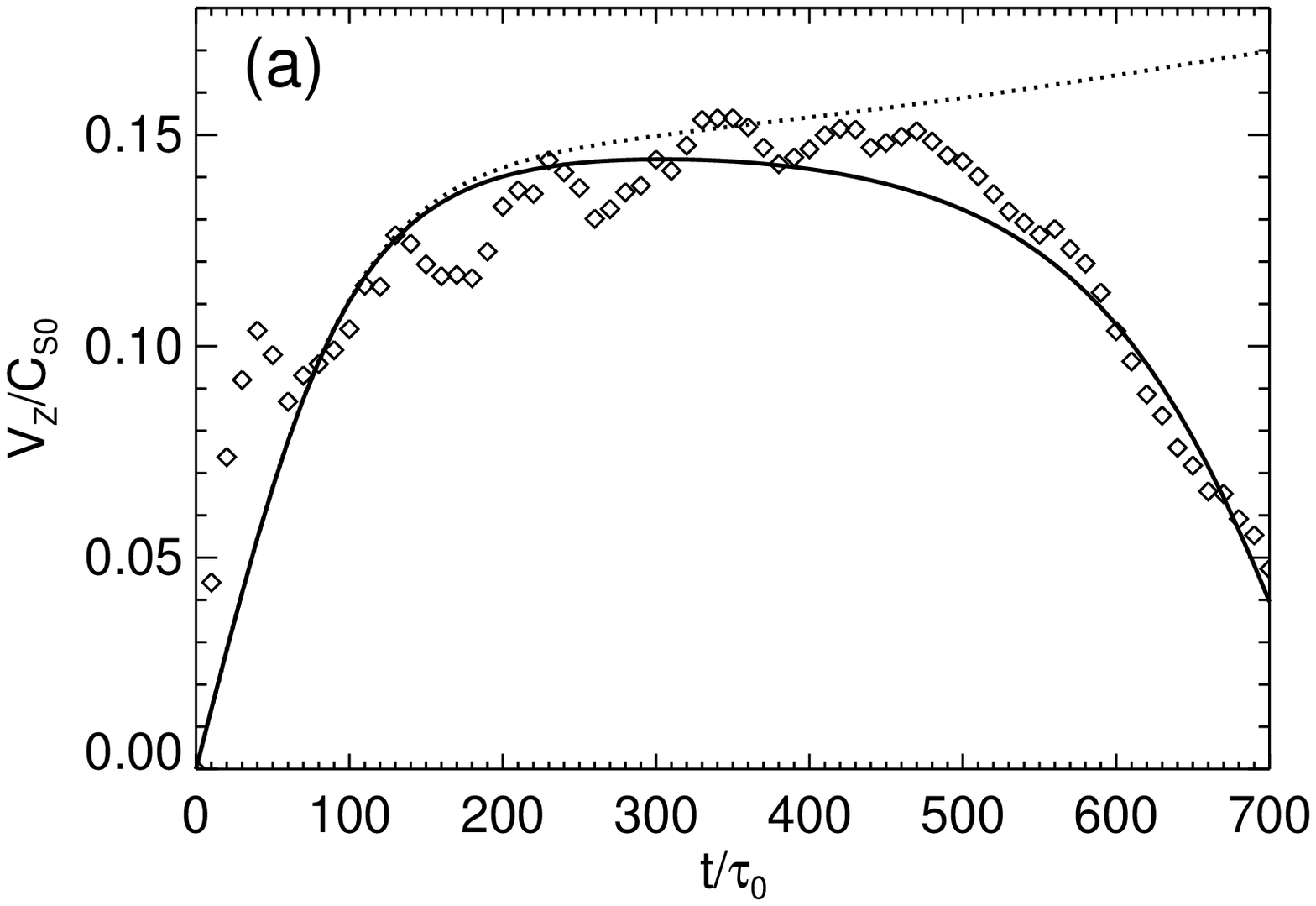}
\label{fig:vz_tft}}\\
\subfigure{\includegraphics[clip,scale=0.5]{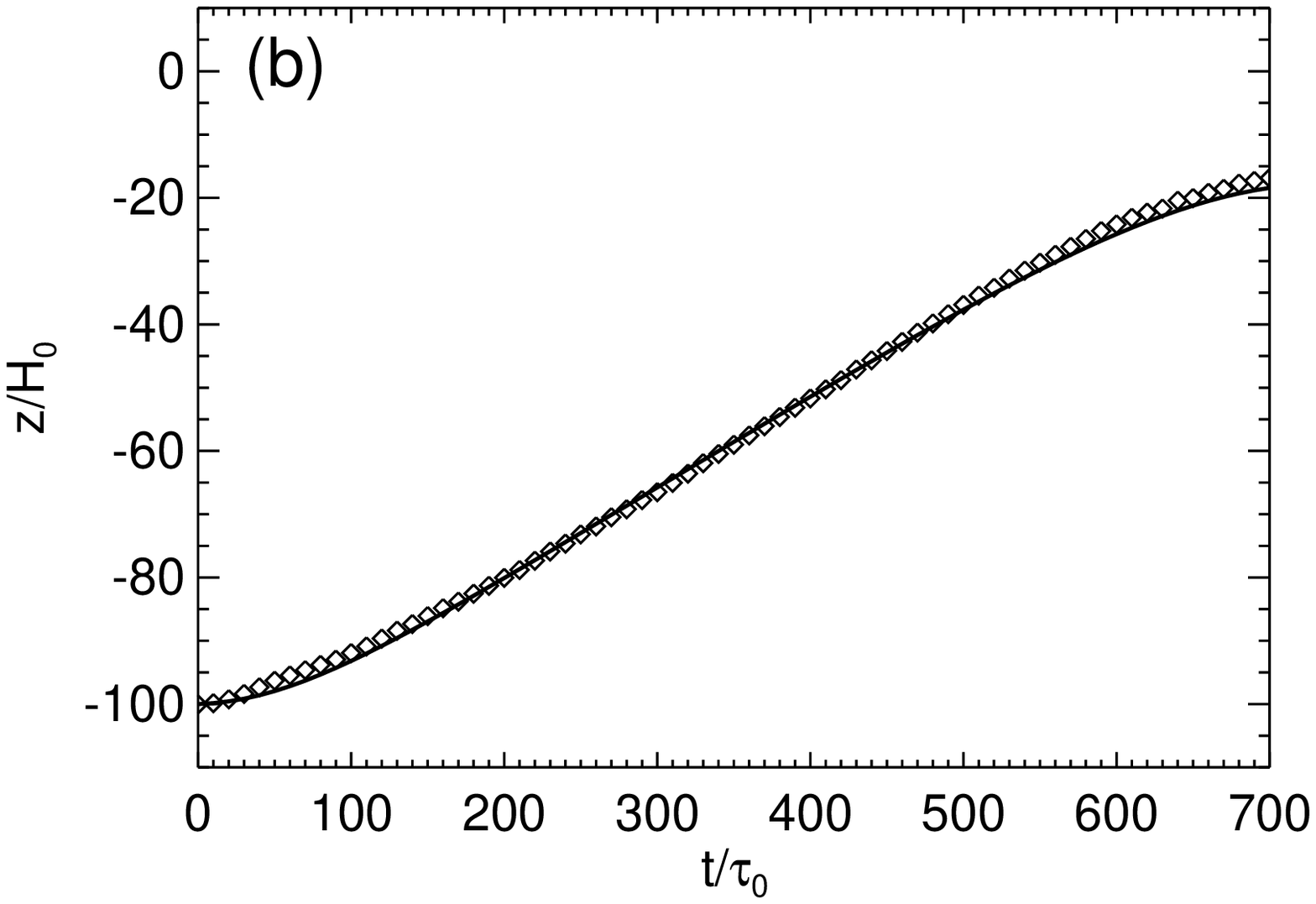}
\label{fig:z_tft}}
\caption{\subref{fig:vz_tft}: The variation of rise velocity
of the rising flux tube with time.
Diamonds indicate the result of the numerical simulation.
Solid line shows our analytic model,
while dotted line represents the model by
\citet{fan98}.
\subref{fig:z_tft}: Height-time relation
of the numerical results (diamonds)
and the analytic model (solid line).
We use $C_{\rm D}=2.0$ and $F=0.001$.}
\label{fig:tft}
\end{center}
\end{figure}

In Figure \ref{fig:vz_tft},
the variation of the rise speed at the tube center
is presented in comparison with the analytic model
(the dotted line is for the model by \citet{fan98}
using $\Delta\rho$
and the solid line for our model using $\Delta\rho'$),
while the heights of the tube centers
of the numerical and the analytic model
are indicated in Figure \ref{fig:z_tft}.
We use $C_{\rm D}=2.0$ and $F=0.001$.
In Figure \ref{fig:vz_tft},
the solid curve levels off after $t/\tau_{0}=100$.
This means that the wake develops from this time
and the drag force becomes more efficient.
\citet{par75} analytically calculated
the terminal velocity of the rising.
Considering Equation (\ref{eq:eqm1}) $=0$
and using Equations (\ref{eq:ini1}) and (\ref{eq:ini2}),
the terminal velocity can be obtained as
\begin{eqnarray}
  V_{z\,\rm term}
  = \left[
      -\frac{\Delta\rho}{\rho} g
      \frac{\pi R}{C_{\rm D}}
    \right]^{1/2}
  = 0.14\, C_{s0}\ ,
\end{eqnarray}
which is rather a good estimation.
In the earlier phase ($t/\tau_{0}<400$),
the torsional oscillation
due to the differential buoyancy
is seen in Figure \ref{fig:vz_tft},
which was mentioned in \S \ref{sec:conv}.

After $t/\tau_{0}=500$,
the rising flux tube decelerates
as the mass on the tube is compressed and piles up,
since the mass cannot persist through
the convectively stable surface.
This deceleration indicates that,
before $t/\tau_{0}=500$,
the cross-sectional evolution
as a whole
can be regarded as radial,
although the tube suffers
aerodynamic deformation;
after that time, however,
the mass pile-up becomes much more efficient so that
the tube decelerates and 
the apex expands horizontally
to become $\nabla$-shaped
(see the solid and the dotted line
in Figure \ref{fig:vz_tft}).

\begin{figure}
\begin{center}
\includegraphics[clip,scale=0.75]{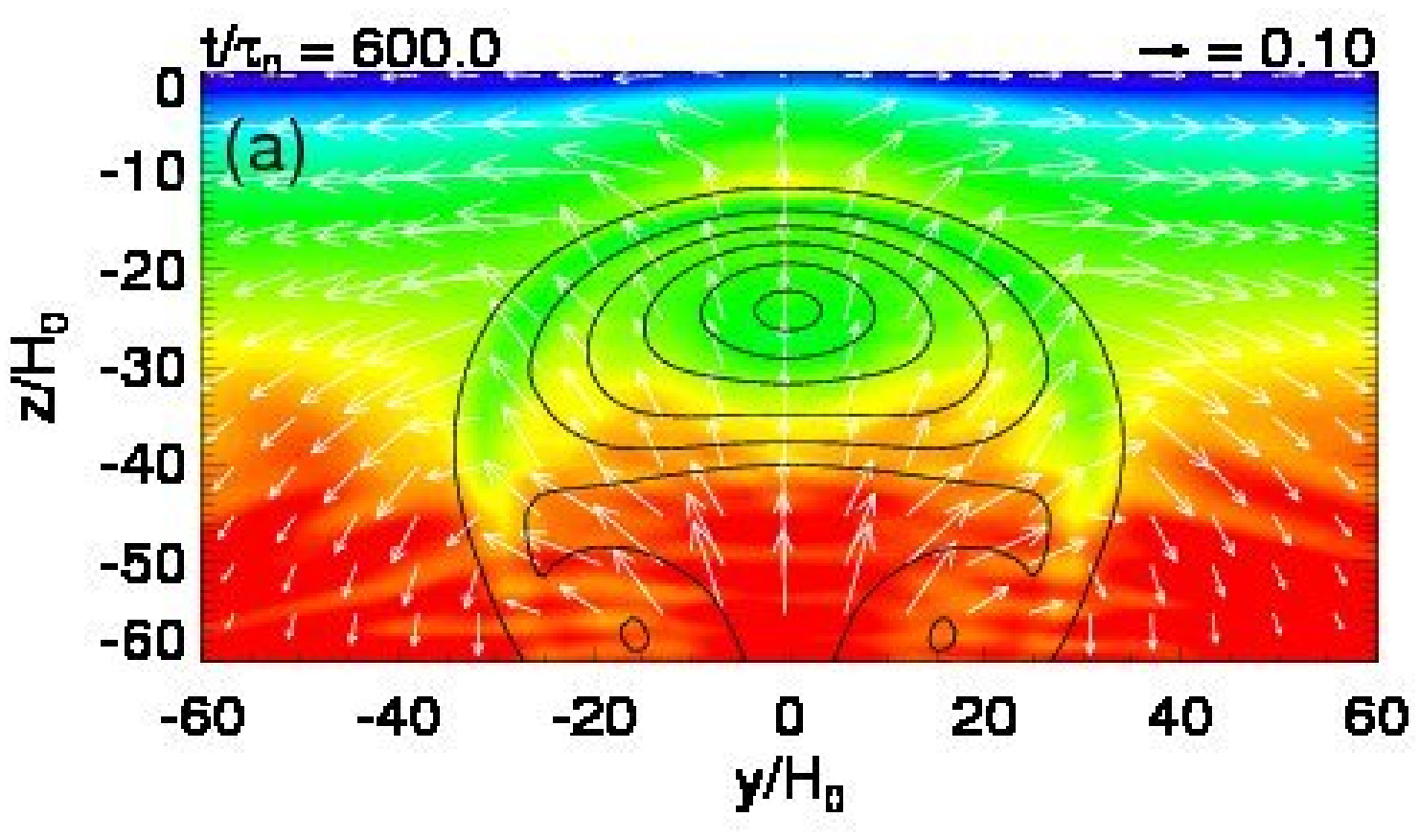}\\
\includegraphics[clip,scale=0.6,angle=-90.]{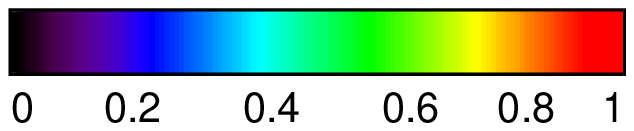}\\
\includegraphics[clip,scale=0.7]{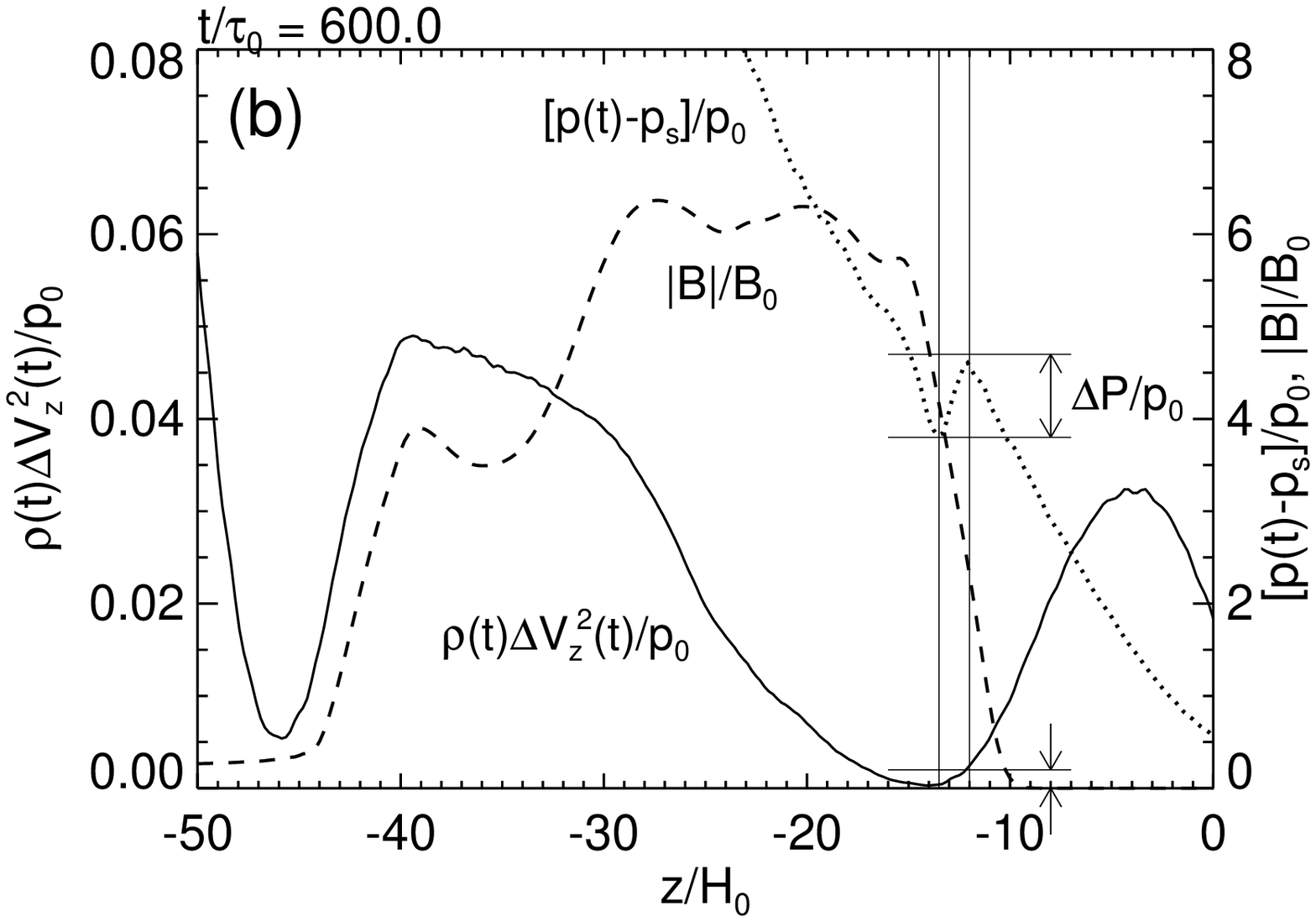}
\caption{
(a): The density accumulation $[\rho (t)-\rho_{\rm s}]/\rho_{0}$,
where $\rho_{\rm s}$ is the background density profile,
and the field lines (contours)
with velocity vectors (white arrows)
at the time $t/\tau_{0}=600$
are shown.
As the tube rises,
the mass piles up
in front of the flux tube
around $(y/H_{0},z/H_{0})=(0,-12)$.
(b): Vertical profiles
of the ram pressure $\rho(t)\Delta V_{z}^{2}(t)/p_{0}$
(solid line),
the pressure excess $[p(t)-p_{\rm s}]/p_{0}$
(dotted line),
and the total field strength $|B|/B_{0}$
(dashed line)
along the symmetric axis $y/H_{0}=0$.
The hump of the pressure excess
$\Delta P/p_{0}$
is indicated by an arrow
in the middle of the figure,
while the corresponding ram pressure
is indicated by
arrows at the bottom
(see text for details).
}
\label{fig:dro}
\end{center}
\end{figure}

Figure \ref{fig:dro}(a) shows the density accumulation
$[\rho (t)-\rho_{\rm s}]/\rho_{0}$,
where $\rho_{\rm s}$ is
the initial background density profile
(see Equation (\ref{eq:pre})),
and the field lines with velocity vectors
at the time $t/\tau_{0}=600$.
It reveals that,
because of the relative mass draining
from the apex of the tube to the flanks,
the density piles up
in front of the rising tube
around $(y/H_{0},z/H_{0})=(0,-12)$
to form a boundary layer
with a finite width ($\sim 40H_{0}$).
In the imaginary sheet case, however,
the fluid would piles up
on the rising sheet
as a boundary layer
with an infinite width,
since there is no draining
(see Paper I).
Therefore,
the factor $F$ (the ratio of the actual accumulation
to the imaginary one) becomes relatively small.

To show that the contribution
of ram pressure
to the pressure excess
(which is related to density excess)
at the tube front
is small enough,
in Figure \ref{fig:dro}(b)
we plot vertical distributions
of ram pressure $\rho(t)\Delta V_{z}^{2}(t)/p_{0}$,
where $\Delta V_{z}(t)=V_{z}(t)-V_{z}(t,z=-14H_{0})$
is the flow velocity relative to the rising tube,
pressure excess $[p(t)-p_{s}]/p_{0}$,
and the total field strength $|B|/B_{0}$
at $t/\tau_{0}=600$
along the symmetric axis $y/H_{0}=0$.
In front of the rising tube ($z/H_{0}=-12$),
pressure excess reveals a hump $\Delta P/p_{0}$,
which is indicated by an arrow
in the middle of Figure \ref{fig:dro}(b).
The corresponding ram pressure
$\rho(t)\Delta V_{z}^{2}(t)/p_{0}$
is also indicated
with arrows in the bottom of the figure.
From this figure,
we can see that
the effect of the ram pressure
$\rho(t)\Delta V_{z}^{2}(t)/p_{0}\sim 0.002$
is much smaller than
the hump of the pressure excess
$\Delta P/p_{0}\sim 1$.
That is,
the contribution of the ram pressure
to the density accumulation
in front of the tube
is negligible.

From Figure \ref{fig:dro}(a),
we can estimate the order of $F$.
The actual mass pile-up along the symmetric axis
$y/H_{0}=0$ is
\begin{eqnarray}
  \Delta\rho_{\rm sim}
  &=& \frac{1}{z_{\rm ph}-z_{\rm apex}(600\tau_{0})}
   \int_{z_{\rm apex}(600\tau_{0})}^{z_{\rm ph}}
   [\rho(600\tau_{0})-\rho_{\rm s}]\, d\zeta \nonumber \\
  &=& 0.73\rho_{0},
  \label{eq:rhosim}
\end{eqnarray}
where $z_{\rm apex}(600\tau_{0})=-12H_{0}$
is the height of the apex of the tube at $t=600\tau_{0}$
(see Figure \ref{fig:dro}).
On the other hand,
the density enhancement
in front of the imaginary emerging sheet at this time
can be estimated from the background profile
$\rho_{\rm s}(z)$ as
\begin{eqnarray}
  \Delta\rho_{\rm acm}&=&\bar{\rho}_{\rm acm}(0)
                       -\bar{\rho}_{\rm acm}(600\tau_{0}) \nonumber \\
  &=& \frac{1}{z_{\rm ph}-z(0)}
      \int_{z(0)}^{z_{\rm ph}}\rho_{\rm s}(\zeta)\, d\zeta
     -\frac{1}{z_{\rm ph}-z(600\tau_{0})}
      \int_{z(600\tau_{0})}^{z_{\rm ph}}
      \rho_{\rm s}(\zeta)\, d\zeta \nonumber \\
  &=& 96.1\rho_{0},
\end{eqnarray}
where $z(600\tau_{0})=-25H_{0}$
is the height of the tube's center
at $t=600\tau_{0}$
(see Figure \ref{fig:z_tft}).
Therefore, we can see that the order of the factor $F$ is
\begin{eqnarray}
  F=\frac{\Delta\rho_{\rm sim}}{\Delta\rho_{\rm acm}}
      =\frac{0.73\rho_{0}}{96.1\rho_{0}}
      =0.007
      =O(10^{-2}).
\end{eqnarray}
Here, we consider the density accumulation
only along the axis $y/H_{0}=0$
in Equation (\ref{eq:rhosim}).
The factor $F$ would be of order $10^{-3}$
if we take it into account
that the accumulation in the neighboring region
is less than that along $y/H_{0}=0$
due to mass draining.

By varying $F$
to fit the analytic model to the obtained data
and plotting the density pile-up in the interior,
we can conclude that
the deceleration of the tube
is caused by the accumulation
of the plasma
ahead of the tube.
\citet{fan98}
and \citet{che06}
also reported this deceleration.
However, their calculations did not include
the convectively stable photosphere
but assumed the non-penetrating or closed top boundaries,
which are reasonable approximations
for the photosphere
compared to our simulation.

\subsection{Further Evolution to the Corona
\label{sec:corona}}
\begin{figure}
\begin{center}
\subfigure{\includegraphics[clip,scale=0.8]{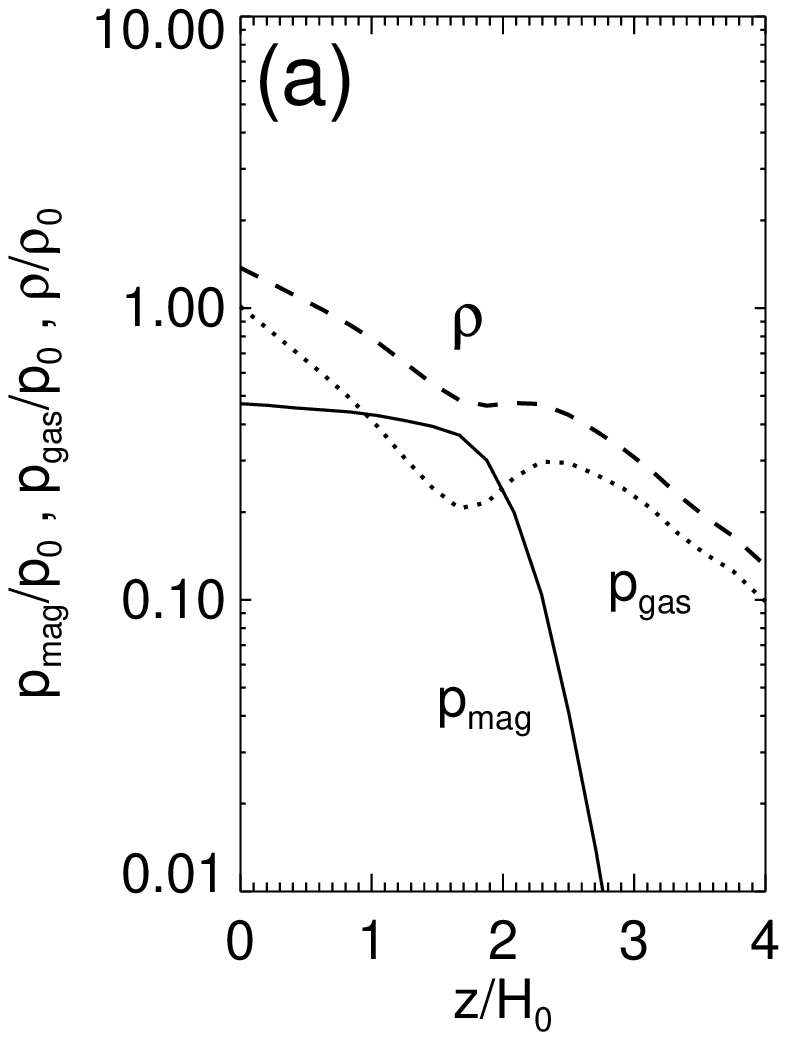}
\label{fig:surface}}\\
\subfigure{\includegraphics[clip,scale=0.8]{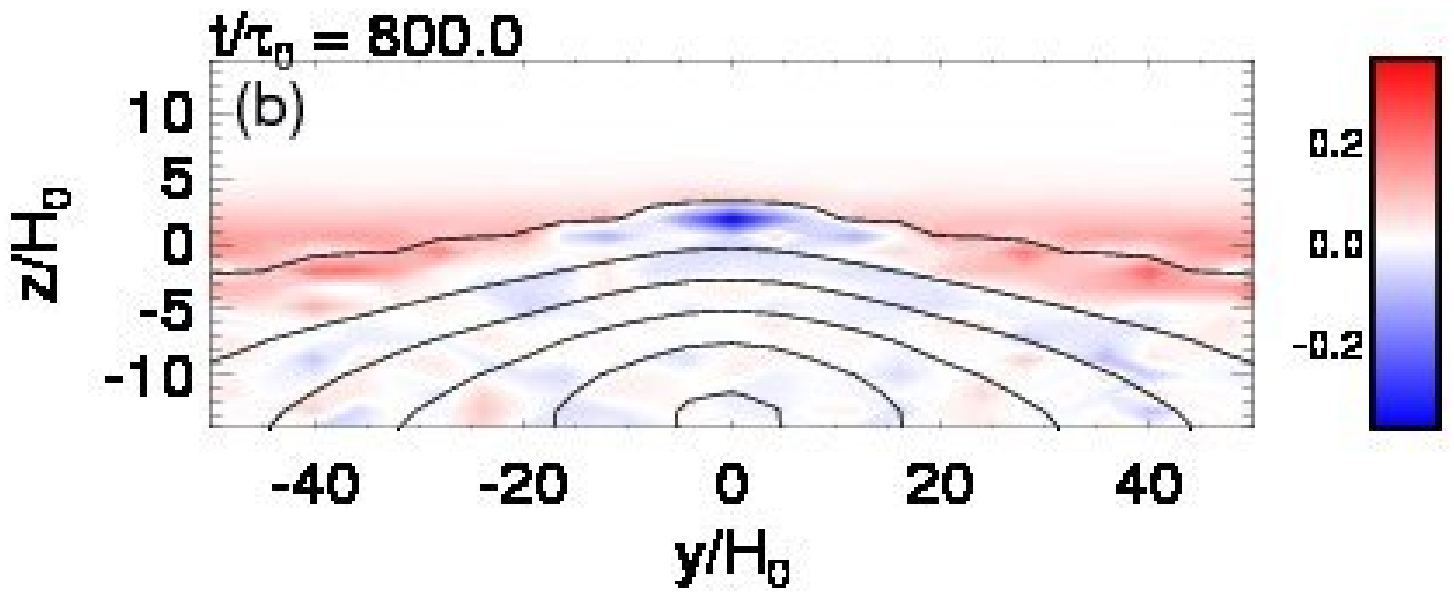}
\label{fig:newcomb}}\\
\caption{(a): Vertical distributions
of the magnetic pressure (solid line),
gas pressure (dotted line), and gas density (dashed line)
along the axis $y/H_{0}=0$ at $t/\tau_{0}=800$.
(b):
Two-dimensional map of the index
$\psi=-\partial\rho/\partial z-\rho^{2}g_{0}/(\gamma p)$.
Color contour indicates $\psi$,
while azimuthal magnetic field lines
are overplotted with solid lines.
The area where index $\psi<0$
is subject to the magnetic buoyancy instability.
}
\label{fig:further}
\end{center}
\end{figure}

After $t/\tau_{0}=800$,
the second-step emergence
from the surface to the corona occurs
due to the magnetic buoyancy instability
(Figures \ref{fig:ro}(f)-(i)).
Figure \ref{fig:surface} shows
the vertical distribution of
magnetic and gas pressure, and density
along the symmetric axis $y/H_{0}=0$
at $t/\tau_{0}=800$,
i.e., just before the secondary evolution starts.
As can be seen from this figure,
the site of the second-step evolution
has a top heavy structure,
and there is a relative ``pressure hill,''
which is consistent with preceding studies
\citep{mag01,arc04}.

To confirm that further evolution
is caused by the magnetic buoyancy instability,
we conduct the same analysis
as in Paper I.
\citet{new61}
revealed the criterion
for the magnetic buoyancy instability
of a flat magnetized atmosphere is
\begin{eqnarray}
  -\frac{\partial\rho}{\partial z}
  < \frac{\rho^{2}g_{0}}{\gamma p}\ .
  \label{eq:newcomb}
\end{eqnarray}
From this relation,
we define the index
\begin{eqnarray}
  \psi = -\frac{\partial\rho}{\partial z}
   - \frac{\rho^{2}g_{0}}{\gamma p}\ ,
\end{eqnarray}
where the area with negative $\psi$
is magnetic buoyancy-unstable.
We show the $\psi$ distribution
in Figure \ref{fig:newcomb}.
As can be seen from Figure \ref{fig:newcomb},
the site of the second-step emergence
($y/H_{0}=z/H_{0}=0$) is $\psi <0$,
i.e., the criterion (\ref{eq:newcomb})
for the instability is satisfied.
Thus, we can conclude
that the further evolution to the corona
occurs because of the magnetic buoyancy instability.
At this site,
the magnetic field intensity amounts to $\sim$1 kG,
and the plasma beta is $\beta\sim 2$.

\begin{figure}
\begin{center}
\subfigure{\includegraphics[clip,scale=0.4]{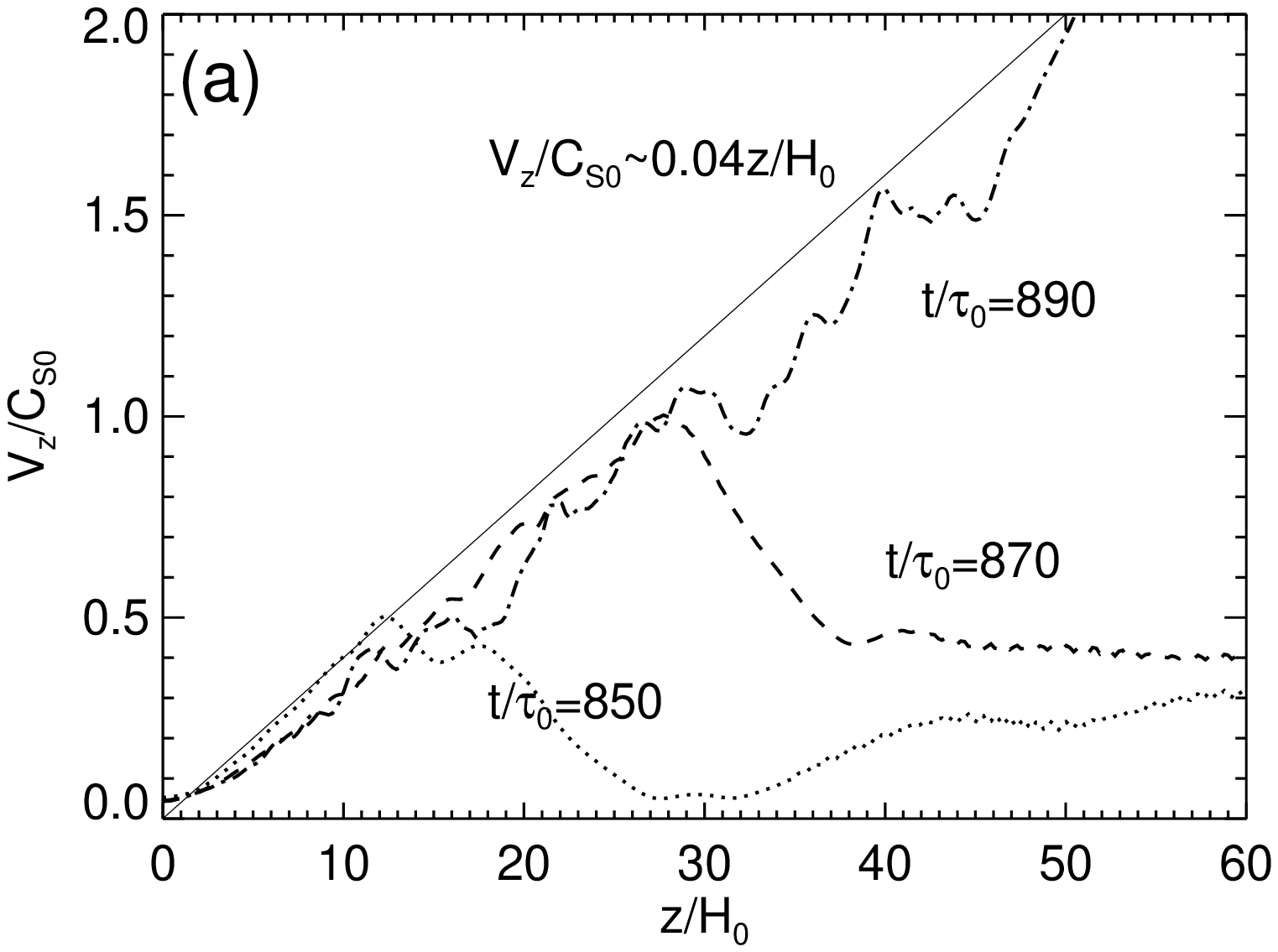}
\label{fig:nonlinear-a}}\\
\subfigure{\includegraphics[clip,scale=0.4]{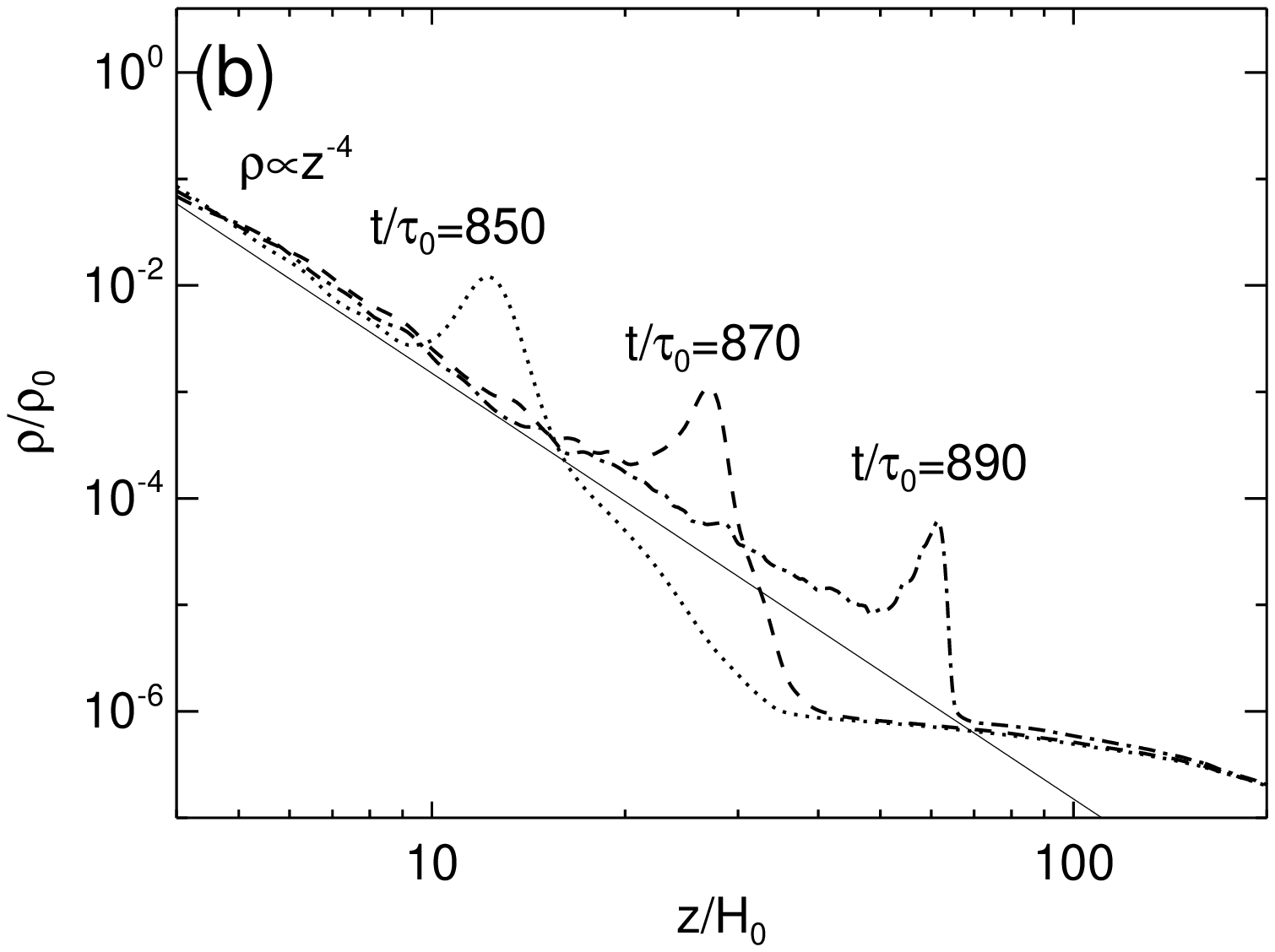}
\label{fig:nonlinear-b}}\\
\subfigure{\includegraphics[clip,scale=0.4]{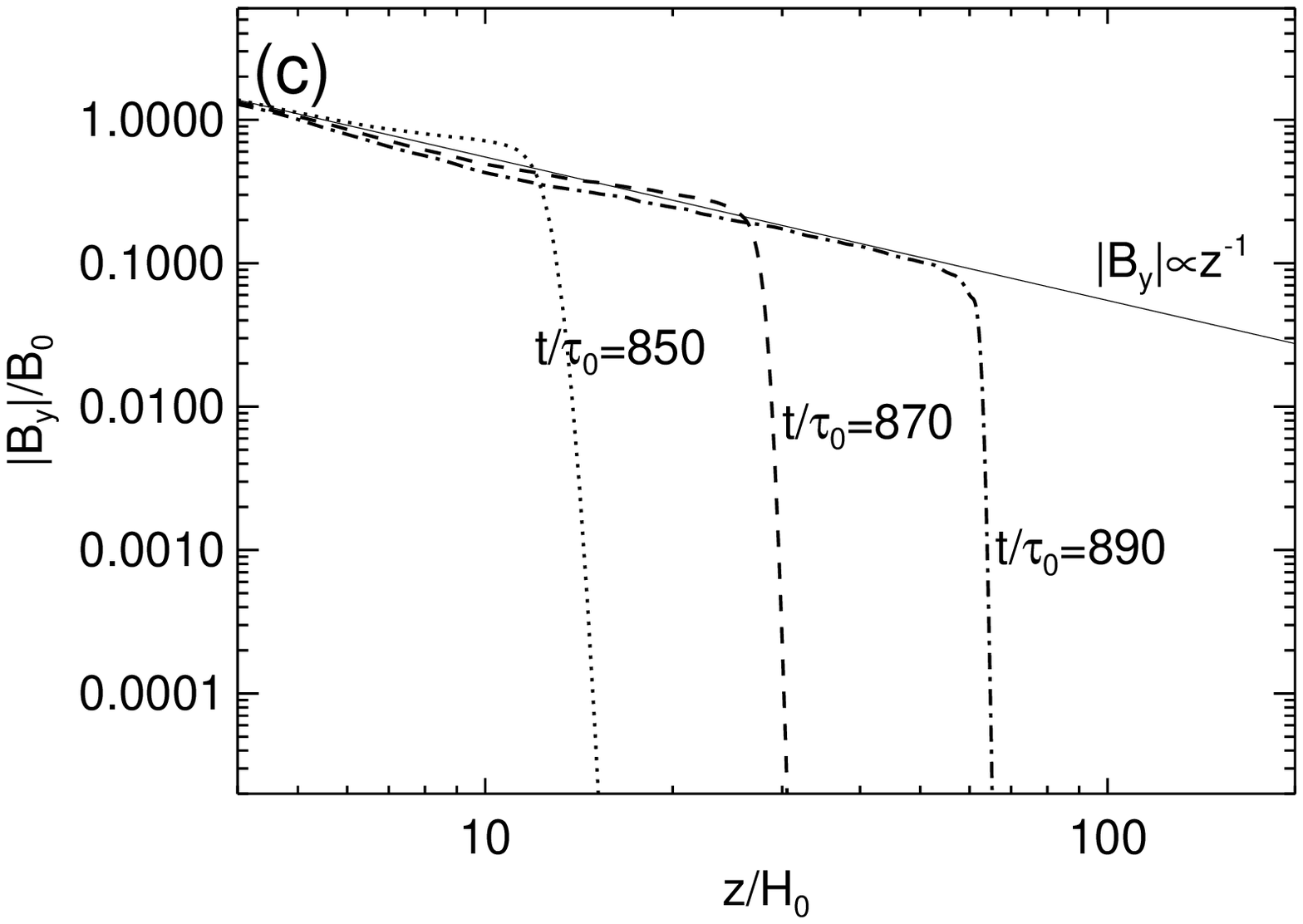}
\label{fig:nonlinear-c}}\\
\caption{(a): Distribution of the upward velocity
along the vertical axis $y/H_{0}=0$.
Dotted, dashed, and dash-dotted lines indicate
the distribution at $t/\tau_{0}=850$, 870,
and 890, respectively.
The solid line shows the theoretical velocity-height relation
according to \citet{shi89}.
(b): Distribution of the gas density.
(c): Distribution of the horizontal component
of the magnetic field.
}
\label{fig:nonlinear}
\end{center}
\end{figure}

In the upper atmosphere above the surface,
the azimuthal component of the magnetic field
is very dominant.
The second-step nonlinear evolution develops
in a self-similar way;
the expansion law was given
by \citet{shi89}.
It can be described as follows:
\begin{eqnarray}
  V_{z}/C_{s0} = az/H_{0}\ ,
\end{eqnarray}
\begin{eqnarray}
  \rho\propto z^{-4}\ ,
\end{eqnarray}
and
\begin{eqnarray}
  |B_{y}|\propto z^{-1}\ ,
\end{eqnarray}
where $a\sim 0.05$
when plasma-$\beta$ is $0.5-2.0$.
The above relations are plotted
in Figure \ref{fig:nonlinear}
for $t/\tau_{0}=850$, 870, and 890.
In Figure \ref{fig:nonlinear-a},
we use $a=0.04$,
which is appropriate
because plasma-$\beta$ was $\sim 2$
at the point of the further evolution
($z\sim 0$ at the time $t/\tau_{0}=800$).

In the final phase, at $t/\tau_{0}=980$,
the coronal loop is formed
with $400H_{0}=80,000\ {\rm km}$ in width
and $200H_{0}=40,000\ {\rm km}$ in height
(see Figure \ref{fig:ro}(i)).
The size of the loop is similar to
that of the case with a flux sheet
(Paper I).
The photospheric field strength
is $400-700\ {\rm G}$ at this time.

\section{Dependence on the Initial Twist
and the Field Strength
\label{sec:param}}

We carry out five additional calculations
to investigate the dependence
of the tube's evolution
on the twist and the field strength.
These runs are summarized in Table \ref{tab:param}.
Parametric surveys on the twist
and the field strength
have done by \citet{mor96}, \citet{emo98},
\citet{mag01}, \citet{mur06}, \citet{mur08},
and \citet{tor11}.
They found that the tube rises faster
as the twist and the field becomes stronger.

\begin{figure}
\epsscale{0.8}
\plotone{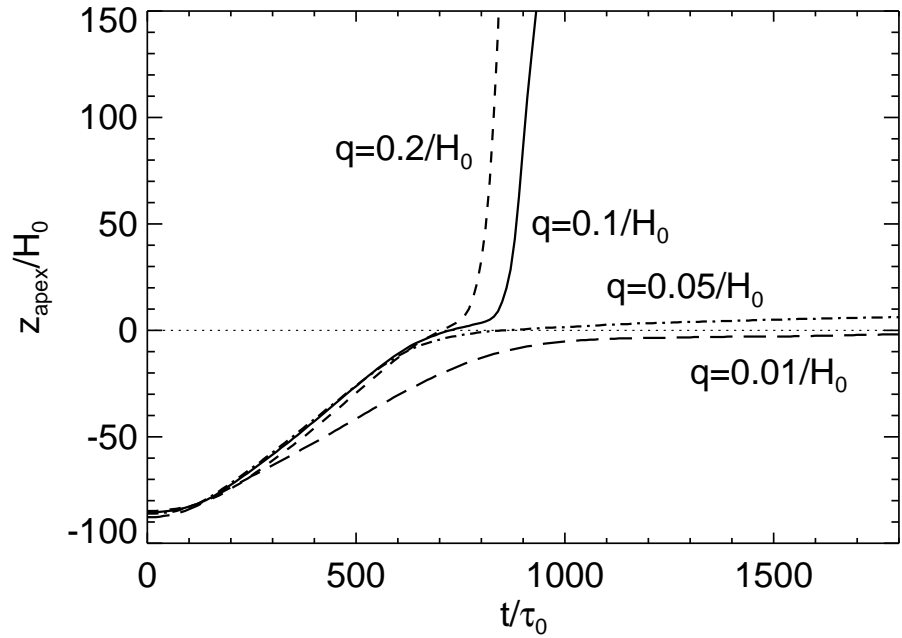}
\caption{Height-time relation
of the top of the flux tube.
Dashed, solid, dash-dotted,
and long dashed lines represent
cases for $q=0.2/H_{0}$,\
$0.1/H_{0}$,\ $0.05/H_{0}$,\ and $0.01/H_{0}$,
respectively.
Dotted line indicates the photospheric level.}
\label{fig:param}
\end{figure}

\begin{figure}
\begin{center}
\includegraphics[clip,scale=0.4]{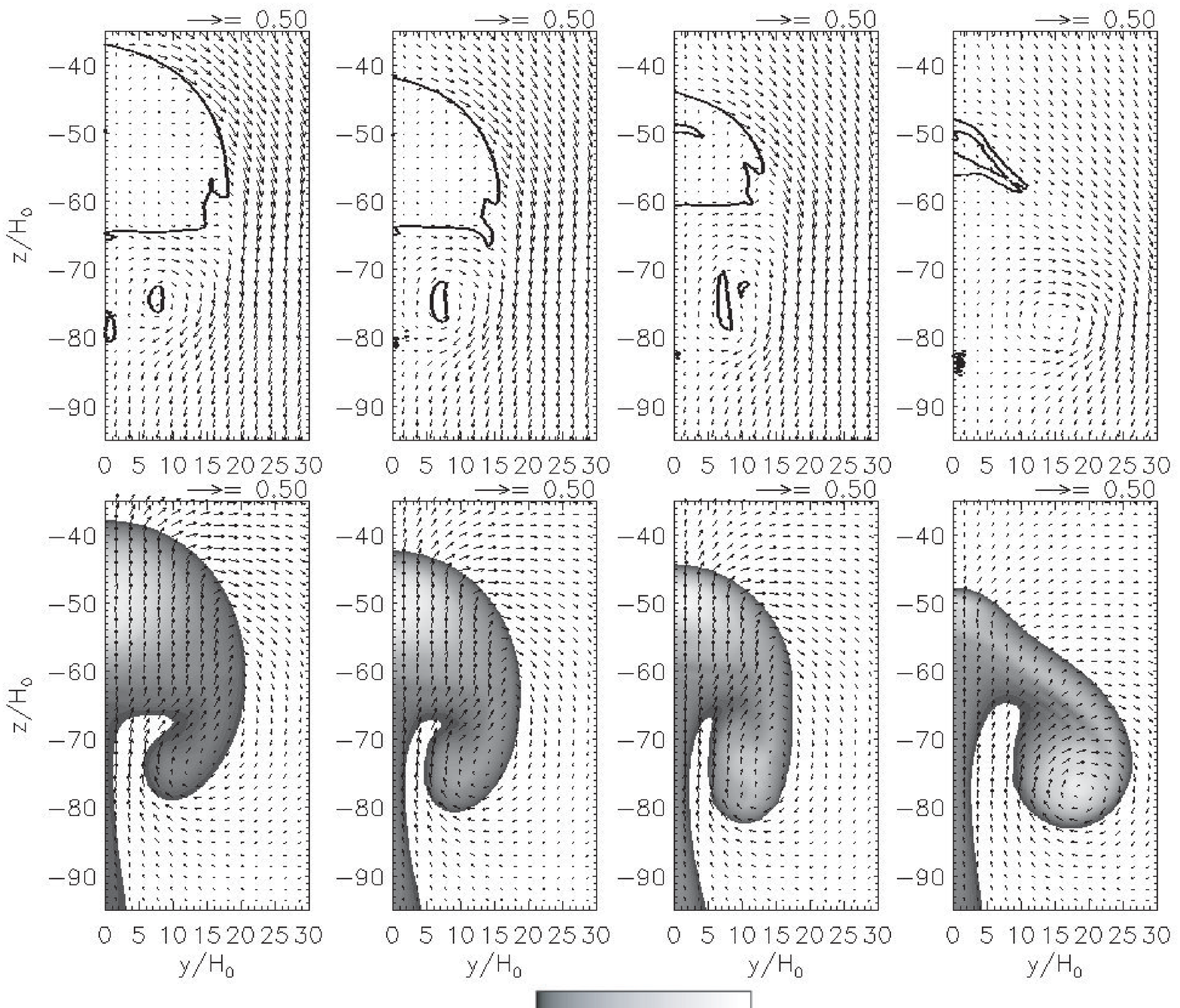}\\
\includegraphics[clip,scale=0.6,angle=-90.]{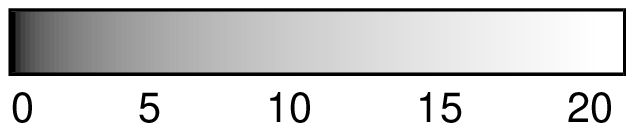}
\caption{Cross-section of the flux tubes
with different initial twist
when their centers are at $z/H_{0}=-50$.
From left to right,
the twist parameters
and their corresponding times are
$q=0.2/H_{0}$,\ $0.1/H_{0}$,\
$0.05/H_{0}$,\ and $0.01/H_{0}$,
and $t/\tau_{0}=460$,\ 410,\ 390,\ and 450,
respectively.
{\it Top:} The velocity relative
to the apex of the tube and the equipartition line
(see text for details).
{\it Bottom:} The longitudinal magnetic field strength
and the flow fields.}
\label{fig:crossec}
\end{center}
\end{figure}

\begin{figure}
\begin{center}
\subfigure{\includegraphics[clip,scale=0.8]{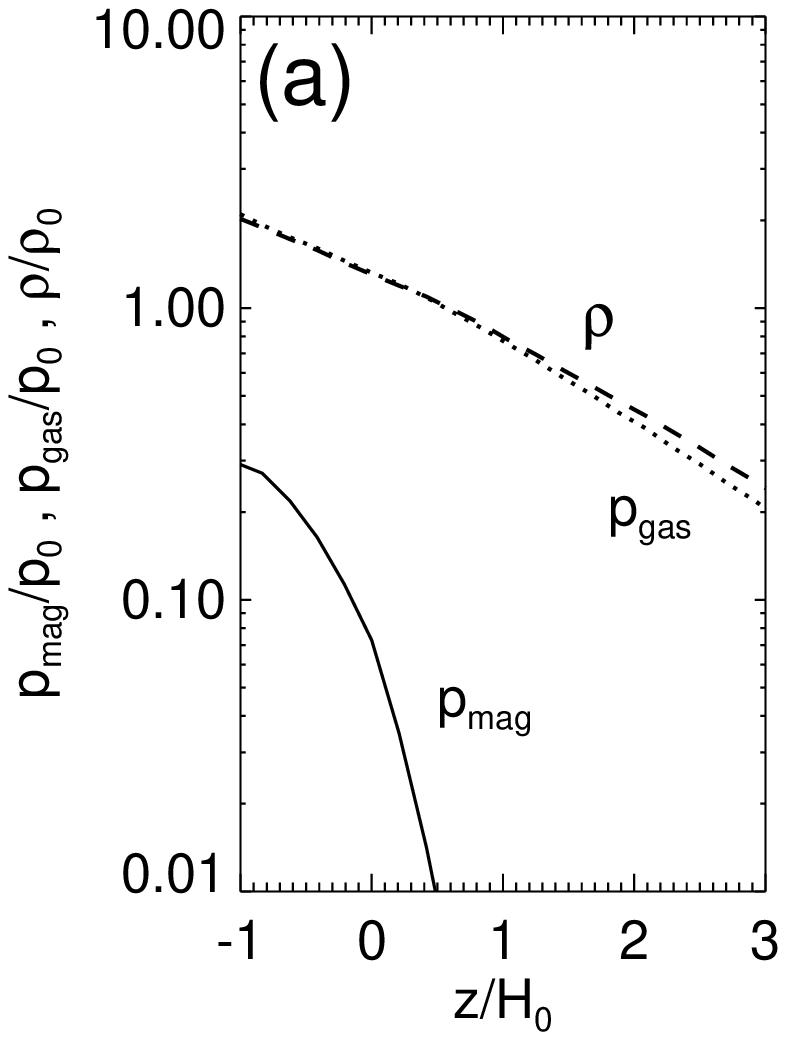}
\label{fig:surface2}}\\
\subfigure{\includegraphics[clip,scale=0.8]{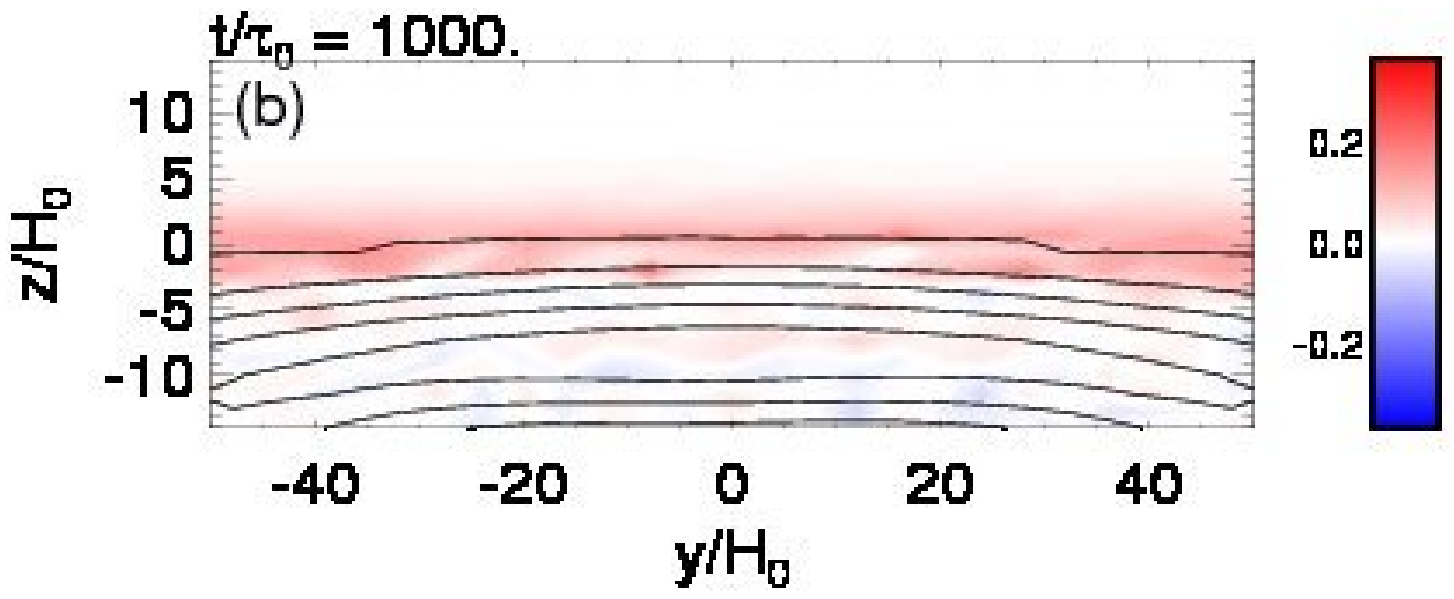}
\label{fig:newcomb2}}\\
\caption{The plots
for the case with $q=0.05/H_{0}$
at $t/\tau_{0}=1000$.
(a): Vertical distributions
of the magnetic pressure (solid line),
gas pressure (dotted line), and gas density (dashed line)
along the axis $y/H_{0}=0$.
(b):
Two-dimensional map of the index
$\psi$ (see \S \ref{sec:corona}).
}
\label{fig:further2}
\end{center}
\end{figure}

Figure \ref{fig:param} shows
the result of the parameter survey
on the tube twist.
In this figure,
height-time relations of the tube tops
are shown.
In cases of a weaker twist
($q=0.01/H_{0}$ and $0.05/H_{0}$),
the flux tube shows
``failed emergence.''
That is, the rising flux cannot
pass through the photosphere.
The criterion for
this failure is $q \lesssim 0.05/H_{0}$.
In Paper I,
we found that,
when the initial field strength
or the total magnetic flux is too weak,
the resulting development also reveals
``failed'' evolution.
Figure \ref{fig:crossec} shows
the flux tubes
with different initial twist
when each tube center
reaches $z/H_{0}=-50$.
From left to right,
the twist parameter decreases.
The upper panels present
the flow field relative to the tube apex,
$\mbox{\boldmath $V$}_{\rm rel}
=\mbox{\boldmath $V$}-\mbox{\boldmath $V$}_{\rm apex}$,
and the corresponding equipartition line.
The equipartition line indicates
the tube boundary
where the kinetic energy density
of the relative flow
$e_{\rm kin}=\rho V_{\rm rel}^2/2$
equals the energy density of the azimuthal field
$e_{\rm mag}=B_{\phi}^2/(8\pi)$.
The lower panels show
the longitudinal field strength $B_{x}$
and the flow field
$\mbox{\boldmath $V$}$.
As can be seen from Figure \ref{fig:crossec},
with a decreasing twist,
the main tube is deformed by the external flow
and the wake develops.
In cases of a weaker twist,
the counter-rotating vortices
contain a large portion
of the magnetic flux.
Thus, when the tube reaches the surface,
the apex of the tube
cannot hold sufficient flux
for the further evolution,
which leads to ``failed emergence.''
The situation can also be explained as follows;
when the initial twist is weak,
the flux tube expands and extends
very widely beneath the surface.
Thus, the magnetic buoyancy falls short
of compressing the flux
to satisfy the criterion (\ref{eq:newcomb})
for the second-step emergence.
In Figure \ref{fig:further2},
we show the distribution near the surface
for the case $q=0.05/H_{0}$
at the time $t/\tau_{0}=1000$.
Figure \ref{fig:surface2} plots
the magnetic pressure,
the gas pressure,
and the density along the symmetric axis
$y/H_{0}=0$.
The subsurface magnetic field
is not fortified enough
so that the magnetic pressure is much smaller
than the plasma pressure.
Therefore, the index $\psi$
(see \S \ref{sec:corona}) is not negative
in almost the whole area
shown in Figure \ref{fig:newcomb2},
which means that the further evolution
to the solar corona
due to the magnetic buoyancy instability
does not take place
in this weaker twist case.

\begin{figure}
\epsscale{0.8}
\plotone{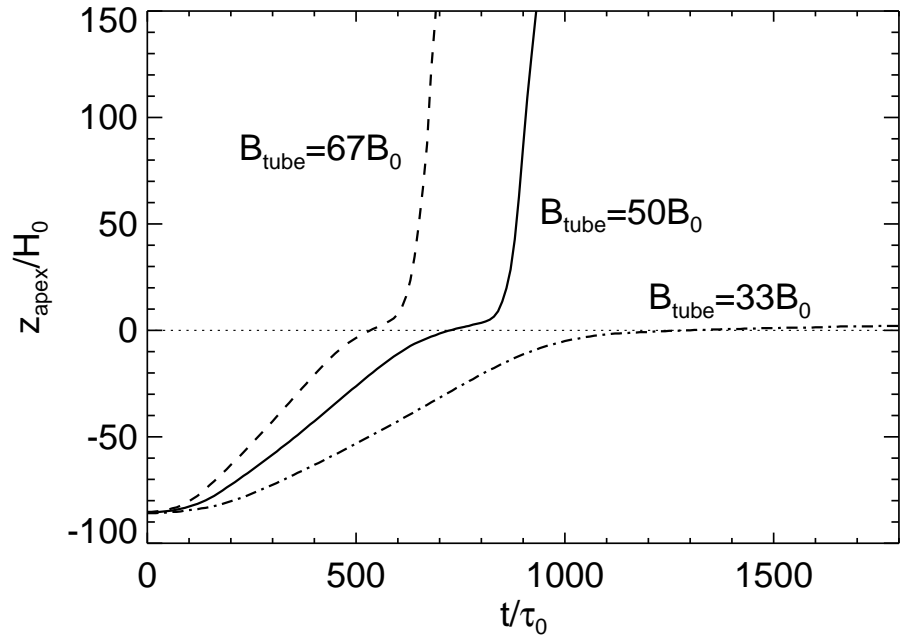}
\caption{Height-time relation
of the top of the flux tube.
Dashed, solid, and dash-dotted lines represent
cases for $B_{\rm tube}=67B_{0}$, $50B_{0}$, and $33B_{0}$,
respectively.
Dotted line indicates the photospheric level.}
\label{fig:btube}
\end{figure}

Figure \ref{fig:btube} shows
the height-time relations
for the cases with
$B_{\rm tube}=67B_{0}$, $50B_{0}$, and $33B_{0}$,
which are indicated with dashed, solid,
and dash-dotted lines.
We can conclude from this figure
that the rise speed is faster with the stronger field.
As is the case with the dependence on the twist,
we can say that,
if the initial field is too weak,
the rise speed is slower
and the flux tube cannot pass through the surface
to rise further into the corona.
The general tendency of the parametric studies
varying the tube's twist and the strength
is consistent with previous
two- and three-dimensional experiments
\citep{mor96,emo98,mag01,mur06,mur08,tor11}.

\section{Summary and Discussion\label{sec:discuss}}

The numerical experiments studied in this paper
are on the cross-sectional evolution
of the emerging twisted flux tube.
In this section,
we summarize the calculations
presented above
and discuss the results.
The discussions are mainly
in connection with Paper I.
We can predict the conditions
for three-dimensional experiments
by considering the results
of the present and previous two-dimensional studies.
So, in this section,
we discuss the two-dimensional results
in connection with our future three-dimensional experiments.

In this paper,
we have studied the dynamical evolution
of the twisted flux tube
that emerged from the deep convection zone.
For the typical case,
the initial flux tube is located
at $z=-100H_{0}=-20,000\ {\rm km}$
with the axial field strength
$B_{\rm tube}=50B_{0}=1.5\times 10^{4}\ {\rm G}$,
the tube radius
$R_{\rm tube}=5H_{0}=1000\ {\rm km}$,
and the total magnetic flux
$\Phi=4.7\times 10^{20}\ {\rm Mx}$.
The tube is initially buoyant
and, thus, begins to rise through
the convection zone.
Halfway to the surface,
after $t=100\tau_{0}=2500\ {\rm s}$,
the periphery of the tube
is peeled away to develop a wake,
composed of vortex rolls
and a long-drawn tail,
due to the external flow
from the apex to the flanks.
As a result,
the aerodynamic drag
becomes more effective.
However, the azimuthal field
prevents the flux tube
from being fragmented into 
a pair of counter-rotating vortices
moving away from each other,
so that the expansion as a whole
can be said to be radial.
The picture of the emergence
within the convective layer
is similar to previous papers
\citep[e.g.][]{mor96}.

The essential difference
from these preceding studies
is the effect of
the convectively stable photosphere;
the apex of the tube expanding horizontally
and the rising motion turning into the deceleration
after $t=500\tau_{0}=1.3\times 10^{4}\ {\rm s}$,
because the rising tube comes close
to the photosphere.
The deceleration occurs at ten times deeper
than previously reported by \citet{mag01};
the deceleration depth is $\sim -500\ {\rm km}$
in Magara's case,
while it is $\sim -5000\ {\rm km}$ in our case,
which shows that the flux tube slows down
before the tube itself reaches the surface.
There seems to be
another mechanism for the slowdown
other than the tube itself entering the photosphere.
The plasma on the flux tube piles up
between the flux and the convectively stable surface,
and depresses the flux below.
We confirmed this effect
by comparing the numerical result
with an extended analytic model
(see Section \ref{sec:analytic}).
As a result,
the $\nabla$-shaped emerging flux extends widely
beneath the surface
($\sim 40,000\ {\rm km}$ in width).
In this stage,
the azimuthal field component is dominant,
since the twist of the flux tube increases
as the tube rises.
When the horizontal flux at the photosphere
becomes unstable to the magnetic buoyancy instability
at $t=800\tau_{0}=2.0\times 10^{4}\ {\rm s} $,
the second-step emergence to the corona
takes place (``two-step emergence'').
The evolution above the surface
is similar to that of \citet{mag01}.
The nonlinear evolution is
well described by the expansion law
given by \citet{shi89}.
The consequent coronal loop
at $t=980\tau_{0}=2.5\times 10^{4}\ {\rm s}$
has its width $\sim 400H_{0}=80,000\ {\rm km}$
and height $\sim 200H_{0}=40,000\ {\rm km}$,
and the photospheric field is
$(1.3-2.3)B_{0}=400-700\ {\rm G}$.

By performing parameter studies,
we found that the initial twist
necessary for the further evolution
is $q>0.05/H_{0}=2.5\times 10^{-4}\ {\rm km}^{-1}$
when the axial field
$B_{\rm tube}=50B_{0}=1.5\times 10^{4}$ G.
If $q$ is less than this value,
the emerging tube remains
within the convection zone
(``failed emergence'').
We also found that
the flux with a weaker field strength
shows the failed emergence.
When the tube twist is
$q=0.1/H_{0}=5.0\times 10^{-4}\ {\rm km}^{-1}$,
the flux with
$B_{\rm tube}=33B_{0}=1.0\times 10^{4}$ G
cannot rise further above the surface.

In Paper I,
we carried out numerical experiments
on the two-dimensional undular evolution
of the flux sheet
from the deep convection zone
($z=-20,000\ {\rm km}$).
We found that
the emerging flux sheet also decelerates
in the convection zone.
However, deceleration depth is
deeper than that of the present study.
Table \ref{tab:discuss}
summarizes the characteristic values
of both types of emergence.
As seen from Table \ref{tab:discuss},
the initial conditions
(the field strength,
the total magnetic flux,
the sheet thickness and the tube radius,
and the depth)
are similar to each other,
while the obtained values
concerning the evolutions
in the convection zone
(the arrival time at the photosphere
and the deceleration depth)
are different.
That is, the twisted flux tube
rises faster than the flux sheet
and the deceleration occurs at a higher altitude.
The difference between the two cases comes from
the geometry of the emerging fluxes,
i.e., a sheet or a tube.
The plasma on the emerging flux tube
can flow around the tube's cross-section
so that the tube rises faster.
The fluid on the flux sheet,
however,
drains only along the magnetic field lines
to both troughs,
so that the plasma suppresses and slows down
the rising sheet below
halfway to the surface.
As for the twisted flux tube,
the deceleration occurs in the later phase
when the tube itself approaches the surface
and begins to expand horizontally
(see \S \ref{sec:conv} and \S \ref{sec:analytic}).
It should be noted that,
in the initial state,
the flux sheet is in hydrostatic equilibrium
while the flux tube
is in mechanically non-equilibrium.
In addition,
the mass draining
along the longitudinal field line
is not considered
in the present tube calculation.
Three-dimensional experiments
are required to investigate
these effects.

By conducting the parameter survey,
we found that
the emerging flux tube cannot rise further
if the initial twist is too weak
(``failed'' case),
because the weak azimuthal field cannot hold
the tube's coherency
and thus cannot hold the field intensity
necessary for the second-step emergence.
Such a flux remains below the surface
and floats around the solar interior.
The condition
for the ``two-step emerging'' flux tube
that yields a realistic active region
at the surface
is $B_{\rm tube}\gtrsim 50B_{0}=1.5\times 10^{4}\ {\rm G}$
with $q\gtrsim 0.1/H_{0}=5.0\times 10^{-4}\ {\rm km}^{-1}$
at $z=-100H_{0}=-20,000\ {\rm km}$.
In Paper I,
it was also found that there is a threshold
of the flux sheet strength
and the total flux
for further emergence.
That is to say,
the emerging flux with a weak twist
also shows ``failed emergence''
as well as the flux with a weak field strength
and an insufficient total flux.
If we put these results
obtained from the two experiments together,
the preferable initial conditions
for the three-dimensional calculation
using the twisted flux tube
from the same depth ($z=-20,000\ {\rm km}$)
are the field strength
$\sim$ a few $10^{4}\ {\rm G}$,
the total magnetic flux of
$\sim 10^{21}-10^{22}\ {\rm Mx}$,
and sufficient twist
($q>0.05/H_{0}=2.5\times 10^{-4}\ {\rm km}^{-1}$).
In future three-dimensional simulations,
we will take into account
the above mentioned conditions.

The comparison with the results
by thin-flux-tube experiments
(see \S \ref{sec:intro})
has been discussed in Paper I.
In Figure 10 of Paper I,
the typical case of the present calculation
(Case 1 with $1.5\times10^{4}\ {\rm G}$
and $4.7\times 10^{20}\ {\rm Mx}$)
locates in the middle of the area
for the two-step emergence.
\citet{mor95} found that
the magnetic flux tube with weaker field
``explodes'' within the interior
and never reaches the surface;
when the pressure gap
between inside and outside the initial flux tube
is too small,
i.e., when the initial magnetic field is too weak,
the tube will be collapsed
at a certain height,
since the pressure gap decreases.
In Paper I,
we found that the cases showing two-step emergence
would have survived the ``explosion''
during their ascents through the interior.
Therefore, we can see that
the typical tube (the present Case 1)
also would have emerged
through the convection zone
without suffering explosion.


%% Included in this acknowledgments section are examples of the
%% AASTeX hypertext markup commands. Use \url without the optional [HREF]
%% argument when you want to print the url directly in the text. Otherwise,
%% use either \url or \anchor, with the HREF as the first argument and the
%% text to be printed in the second.

\acknowledgments
Numerical computations were
carried out on NEC SX-9
at the Center for Computational Astrophysics, CfCA,
of the National Astronomical Observatory of Japan,
and on M System (Fujitsu FX1)
and V System (NEC SX-9)
of JAXA Supercomputer System.
The page charge of this paper
is partly supported by CfCA.
The authors thank Drs. H. Isobe and K. Shibata
of Kyoto University
and Dr. Y. Fan of the High Altitude Observatory,
the National Center for Atmospheric Research.
We are grateful to Mr. David Ward,
the GCOE program instructor
of the University of Tokyo
for proofreading/editing assistance.

\clearpage

%% the file, uncomment the \plotone, \plottwo, and
%% If you need a layout that cannot be achieved with \plotone or
%% \plottwo, you can invoke the graphicx package directly with the
%% \includegraphics command or use \plotfiddle. For more information,
%% please see the tutorial on "Using Electronic Art with AASTeX" in the
%% documentation section at the AASTeX Web site,
%% http://www.journals.uchicago.edu/AAS/AASTeX.
%%
%% The examples below also include sample markup for submission of
%% supplemental electronic materials. As always, be sure to check
%% the instructions to authors for the journal you are submitting to
%% for specific submissions guidelines as they vary from

%% This example uses \plotone to include an EPS file scaled to
%% 80% of its natural size with \epsscale. Its caption
%% has been written to indicate that additional figure parts will be
%% available in the electronic journal.

\clearpage

\clearpage

\clearpage

\clearpage

\clearpage

\clearpage

\clearpage

\clearpage

\clearpage

\clearpage

\clearpage

\clearpage

\clearpage

\begin{deluxetable}{crrr}
\tablecaption{Summary of Cases\label{tab:param}}
\tablewidth{0pt}
\tablehead{
\colhead{Case} &
\colhead{$B_{\rm tube}$ (300 G)\tablenotemark{a}} &
\colhead{$R_{\rm tube}$ (200 km)\tablenotemark{b}} &
\colhead{$q$ (0.005 km$^{-1}$)\tablenotemark{c}}
}
\startdata
 1 & 50 & 5 & 0.10 \\
 2 & 50 & 5 & 0.20 \\
 3 & 50 & 5 & 0.05 \\
 4 & 50 & 5 & 0.01 \\
 5 & 67 & 5 & 0.10 \\
 6 & 33 & 5 & 0.10 \\
\enddata
\tablenotetext{a}{Magnetic field strength at the tube center.}
\tablenotetext{b}{Tube radius.}
\tablenotetext{c}{Twist parameter.}
\end{deluxetable}

\clearpage

\begin{deluxetable}{lcc}
\tablecaption{Comparison of the Two Types
of the Flux Emergence\label{tab:discuss}}
\tablewidth{0pt}
\tablehead{
\colhead{} &
\colhead{Undular Emergence} &
\colhead{Non-equilibrium Emergence}\\
\colhead{Characteristic Variables} &
\colhead{of a Flux Sheet\tablenotemark{a}} &
\colhead{of a Twisted Flux Tube\tablenotemark{b}}
}
\startdata
 Initial Field Strength & $1.0\times 10^{4} {\rm G}$ &
  $1.5\times 10^{4}\ {\rm G}$ \\
 Total Magnetic Flux & $1.0\times 10^{21}\ {\rm Mx}$
  \tablenotemark{c} &
  $4.7\times 10^{20}\ {\rm Mx}$ \\
 Sheet Thickness / Tube Radius &
  $1000\ {\rm km}$ &
  $1000\ {\rm km}$\\
 Initial Depth & $-20,000\ {\rm km}$ &
  $-20,000\ {\rm km}$ \\
 Arrival Time at the Surface &
  $4.9\times 10^{4}\ {\rm s}$ &
  $2.0\times 10^{4}\ {\rm s}$ \\
 Deceleration Depth & $-10,000\ {\rm km}$ &
  $-5000\ {\rm km}$ \\
\enddata
\tablenotetext{a}{Typical case of Paper I.}
\tablenotetext{b}{Typical case of this paper (twist $q=0.1/H_{0}$).}
\tablenotetext{c}{The flux sheet is assumed as a prism with a base
  1000 km $\times$ 10,000 km}
\end{deluxetable}

%% Tables may also be prepared as separate files. See the accompanying
%% sample file table.tex for an example of an external table file.
%% To include an external file in your main document, use the \input
%% command. Uncomment the line below to include table.tex in this
%% sample file. (Note that you will need to comment out the \documentclass,
%% \begin{document}, and \end{document} commands from table.tex if you want
%% to include it in this document.)


%% The following command ends your manuscript. LaTeX will ignore any text


\begin{thebibliography}{}
\bibitem[Abbett \& Fisher(2003)]{abb03}
 Abbett, W. P., \& Fisher, G. H. 2003, \apj, 582, 475
\bibitem[Archontis et al.(2004)]{arc04} Archontis, V.,
Moreno-Insertis, F., Galsgaard, K., Hood, A., \& O'Shea, E.
2004, \aap, 426, 1047
\bibitem[Caligari et al.(1995)]{cal95} Caligari, P.,
Moreno-Insertis, F., \& Sch$\ddot{\rm u}$ssler, M.  1995,
\apj, 441, 886
\bibitem[Cheung et al.(2006)]{che06} Cheung, M. C. M., Moreno-Insertis, F.,
\& Sch$\ddot{\rm u}$ssler, M.
2006, \aap, 451, 303
\bibitem[Cheung et al.(2007)]{che07} Cheung, M. C. M.,
 Sch$\ddot{\rm u}$ssler, M., Moreno-Insertis, F.
 2007, \aap, 467, 703
\bibitem[Cheung et al.(2008)]{che08} Cheung, M. C. M.,
 Sch$\ddot{\rm u}$ssler, M., Tarbell, T. D., \& Title, A. M.
 2008, \apj, 687, 1373
\bibitem[Dorch(1999)]{dor99} Dorch, S. B. F., Archontis, V.,
 \& Nordlund, A. 1999, \aap, 352, L79
\bibitem[Dorch(2007)]{dor07} Dorch, S. B. F.
2007, \aap, 461, 325
\bibitem[D'Silva \& Choudhuri(1993)]{dsi93}
 D'Silva, S. \& Choudhuri, A. R. 1993, \aap, 272, 621
\bibitem[Emonet \& Moreno-Insertis(1998)]{emo98} Emonet, T., \& Moreno-Insertis, F.
1998, \apj, 492, 804
\bibitem[Fan et al.(1993)]{fan93}
 Fan, Y., Fisher, G. H., \& Deluca, E. E. 1993, \apj, 405, 390
\bibitem[Fan et al.(1994)]{fan94}
 Fan, Y., Fisher, G. H., \& McClymont, A. N. 1994, \apj, 436, 907
\bibitem[Fan et al.(1998)]{fan98} Fan, Y., Zweibel, E. G., \& Lantz, S. R.
1998, \apj, 493, 480
\bibitem[Fan(2001)]{fan01} Fan, Y. 2001, \apj, 554, L111
\bibitem[Galsgaard et al.(2007)]{gal07}
 Galsgaard, K., Archontis, V., Moreno-Insertis, F., \& Hood, A. W.
 2007, \apj, 666, 516
\bibitem[Hood et al.(2009)]{hoo09}
 Hood, A. W., Archontis, V., Galsgaard, K, \& Moreno-Insertis, F.
 2009, \aap, 503, 999
\bibitem[Isobe et al.(2005)]{iso05} Isobe, H., Miyagoshi, T.,
 Shibata, K., \& Yokoyama, T. 2005, \nat, 434, 478
\bibitem[Longcope et al.(1996)]{lon96} Longcope, D. W., Fisher, G. H., \& Arendt, S.
1996, \apj, 464, 999
\bibitem[MacTaggart \& Hood(2009)]{mac09}
 MacTaggart, D., \& Hood, A. W. 2009, \aap, 507, 995
\bibitem[Magara(2001)]{mag01} Magara, T. 2001, \apj, 549, 608
\bibitem[Manchester et al.(2004)]{man04}
 Manchester, W., IV, Gombosi, T., DeZeeuw, D. \& Fan, Y.
 2004, \apj, 610, 588
\bibitem[Matsumoto \& Shibata(1992)]{mat92} Matsumoto, R., \& Shibata, K.
 1992, \pasj, 44, 167
\bibitem[Moreno-Insertis et al.(1995)]{mor95} Moreno-Insertis, F.,
Caligari, P., \& Sch$\ddot{\rm u}$ssler, M.  1995, \apj, 452, 894
\bibitem[Moreno-Insertis \& Emonet(1996)]{mor96} Moreno-Insertis, F., \& Emonet, T.
1996, \apj, 472, L53
\bibitem[Moreno-Insertis(2006)]{mor06}
 Moreno-Insertis, F. 2006, ASP Conf. Ser., 354, 183
\bibitem[Murray et al.(2006)]{mur06}
 Murray, M. J., Hood, A. W., Moreno-Insertis, F., Galsgaard, K.,
 \& Archontis, V. 2006, \aap, 460, 909
\bibitem[Murray \& Hood(2008)]{mur08}
 Murray, M. J., \& Hood, A. W. 2008, \aap, 479, 567
\bibitem[Newcomb(1961)]{new61} Newcomb, W. A.
 1961, Phys. Fl., 4, 391
\bibitem[Nozawa et al.(1992)]{noz92} Nozawa, S., Shibata, K.,
 Matsumoto, R., Sterling, A. C., Tajima, T., Uchida, Y.,
 Ferrari, A., \& Rosner, R.  1992, \apjs, 78, 267
\bibitem[Parker(1955)]{par55} Parker, E. N.  1955, \apj, 121, 491
\bibitem[Parker(1974)]{par74} Parker, E. N.  1974, \apj, 191, 245
\bibitem[Parker(1975)]{par75} Parker, E. N.  1975, \apj, 198, 205
\bibitem[Parker(1979)]{par79} Parker, E. N.  1979,
Cosmical magnetic fields: Their origin and their activity
\bibitem[Sch$\ddot{\rm u}$ssler(1979)]{sch79} Sch$\ddot{\rm u}$ssler, M. 1979, \aap, 71, 79
\bibitem[Shibata et al.(1989)]{shi89} Shibata, K., Tajima, T.,
Steinolfson, R. S., \& Matsumoto, R.  1989, \apj, 345, 584
\bibitem[Spruit(1981)]{spr81} Spruit, H. C. 1981, \aap, 98, 155
\bibitem[Toriumi \& Yokoyama(2010)]{tor10} Toriumi, S. \& Yokoyama,
  T. 2010, \apj, 714, 505
\bibitem[Toriumi et al.(2011)]{tor11} Toriumi, S.,
  Miyagoshi, T., Yokoyama, T., Isobe, H., \& Shibata, K.
  2011, \pasj, 63, 407
\bibitem[Tortosa-Andreu \& Moreno-Insertis(2009)]{tor09}
 Tortosa-Andreu, A., \& Moreno-Insertis, F. 2009, \aap, 507, 949
\end{thebibliography}
\end{document}